\UseRawInputEncoding
\documentclass[aps,prd,onecolumn,eqsecnum,amsmath,nofootinbib,preprintnumbers]{revtex4}%

\usepackage{color,graphicx,floatrow,subfigure,xcolor}
\usepackage{amsfonts,amssymb,theorem,mathrsfs,times}
\usepackage{bm}
\usepackage{mathtools}
\usepackage{amsfonts,amssymb,theorem,mathrsfs}
\usepackage{dsfont}
\usepackage{setspace}
\usepackage{amsmath}  
\usepackage{xcolor}
\usepackage[colorlinks,linkcolor=blue,anchorcolor=blue,citecolor=blue]{hyperref}   
\usepackage{ulem}   
\usepackage[english]{babel}
\usepackage{slashed}

{\theorembodyfont{\upshape}
	}
{\theorembodyfont{\upshape}
	}
{\theorembodyfont{\upshape}
	}
{\theorembodyfont{\upshape}
	}
{\theorembodyfont{\upshape}
	}
{\theorembodyfont{\upshape}
	}
\addtocounter{MaxMatrixCols}{10}
\newcommand{\dalm}{\kern1pt\vbox{\hrule height 0.9pt\hbox{\vrule width
			0.9pt\hskip 2.5pt\vbox{\vskip 5.5pt}\hskip 3pt\vrule width
			0.3pt}\hrule height 0.3pt}\kern1pt}

\begin{document}
\thispagestyle{empty}

\title{Thermodynamic law and holography dual of accelerating and rotating black hole in Nariai limit}

\author{ Shu Luo$^{a}$\footnote{e-mail
			address: ls040629@mail.ustc.edu.cn}}

\affiliation{${}^a$School of  Physics ,
 University of Science and Technology of China, Hefei, Anhui 230026,
	China}

\date{\today}

\begin{abstract}
 {In this study, we investigate the thermodynamic law  of accelerating and rotating black hole described by rotating C-metric, as well as holography properties in Nariai limit, which are related to Nariai-CFT and Kerr-CFT correspondence. In order to achieve this goal we define a regularized Komar mass with physical interpretation of varying the horizon area from spinless limit to general case, and derive the frist law based on this construction through covariant phase space formalism. Serving for potential future studies, we also reduce the model to a 2-dimensional JT-type action and discuss some of its properties.}	
\end{abstract}

\maketitle		
\section{Introduction}\label{sec:1}
Black holes, serving as extreme gravity objects, have recently led to many theoretical and phenomenological advances~\cite{cai2025pseudospectrum,cao2025pseudospectrumtimedomainanalysiseft,wu2025stabilitygreybodyfactorhayward,luo2024quasinormalmodespseudospectrumtime,luo2024black}.
With the intensive study on the AdS-CFT correspondence, many properties of the duality between quantum gravity systems and gauged fields have been revealed. A rather similar holography composition is about Kerr-CFT duality, which relates to many familiar spacetime, including those belonging to the general type-D family, Plebanski-Demianski solutions~\cite{Plebanski:1976gy}. A special type of them named C-metric~\cite{Griffiths:2009dfa} is supposed to describe accelerating black holes, and has much relation to other interesting spacetime, including Ernst solution and Melvin solution, etc. While C-metric with negative cosmological constant is often regarded as describing one single black hole driven by cosmological strings, the one in asymptotically flat can be interpreted as two black holes accelerated towards or away from each other. The similar argument that makes sense in Kerr-CFT duality is now found to be true for the extremal case of this type of solution, Nariai limit, as well, which we will call it Nariai-CFT. 

We first meet Nariai geometry in extremal Kerr-de-Sitter spacetime, in which the event horizon coincides with the cosmological horizon. In non-rotating case this can reduce to $\mathrm{dS_{2}}\times S^2$. Due to quantum fluctuations, the Nariai solutions are unstable and, once created, they decay through the quantum tunneling process into a slightly non-extremal dS spacetime~\cite{PhysRevD.54.6312}. Another interesting feature of Nariai solution is the instanton related to quantum decay of the dS space accompanied by the creation of a dS black hole pair. The Nariai limit of non-rotating C-metric was analyzed in detail in~\cite{Dias_2003}, arguing that the geometry of the Nariai C-metric is $\mathrm{dS_{2}}\times \tilde{S}^2$, where $\tilde{S}^2$ is the deformed 2-sphere. In this limit the black hole can also be regarded as in full thermodynamic equilibrium, where temperature calculated at two different horizons now coincide~\cite{Booth_1999}. A special exception that only exsits for nonzero acceleration is ``Nariai" flat spacetime, in which there is an accelerating horizon which plays similar role to the cosmic horizon even when the spacetime is asymptotically flat. However, it would be seen that such a horizon is noncompact and thus the geometry would be $\mathrm{dS_2}\times \mathbb{R}^2$.

Since the pioneer work of~\cite{PhysRevD.80.124008}, many interesting properties of the near-horizon limit of extremal Kerr spacetime (NHEK) have been revealed. It is believed that parallel to the crucial properties of duality between quantum gravity in $\mathrm{AdS_{3}}$ and the 2-D conformal theory (due to pure symmetry considerations)~\cite{Brown:1986nw}, the NHEK spacetime is a fibered product of two-dimensional anti-de Sitter space and two-sphere. The spacetime with a fixed $\theta$ is precisely warped $\mathrm{AdS_{3}}$, with which the deformation of the radius of $S^1$ fiberation over $\mathrm{AdS_{2}}$ can lead to an $SL(2, R)\times U(1)$ isometry group. For Nariai case we only need to replace $\mathrm{AdS_{3}(AdS_{2})}$ by $\mathrm{dS_{3}(dS_{2})}$, and the isometry group does not change~\cite{Sakti_2024}. At the same time, consistent boundary conditions can select an asymptotical symmetry group (ASG) that is precisely the same group~\cite{Bredberg_2011}.  Moreover, it was found that there is ``hidden symmetry" of scalar wavefunction in both ``near" and ``far" region, in which the operator can serve as a quadric $SL(2,R)$ Casimir operator~\cite{Castro_2010}. As shown in the following sections, this argument can be naturally extend to spacetime with similar conditions and even slightly different extremal geometry, such as in this work, where the correspondence for ``Nariai" flat C-metric is also verified.

It seems that the near-horizon limit of ``Nariai" flat C-metric spacetime is similar to usual near-extremal ones, but there are something quite different. Asymptotically, the rotating Nariai spacetime is naturally foliated by a timelike “radial” coordinate and the foliations are spacelike. The similarity of Nariai-CFT dual and dS-CFT dual leads to the fact that the dual conformal field theory (CFT) is
expected to live on a space-like surface and the time coordinate emerges from a Euclidean CFT. Such CFTs turn
out to be non-unitary, being exotic compared with standard examples of CFTs~\cite{Doi_2023}. Still, according to the traditional interpretation, if the CFT dual to dS spacetime is supposed to live on the sapcelike boundary $\mathcal{I}_{+}$, then there is the problem of how to interpret the observer and observables, including how to perform asymptotically precise measurements~\cite{witten2001quantumgravitysitterspace}. Yet recently, in lower dimensional $\mathrm{dS_{2}}$ gravity, the systematic theory of holography has been built~\cite{maldacena2020dimensionalnearlysittergravity}, in which the computation of the no-boundary wavefunction of the universe is essentially identical to the computation of the partition function for the euclidean $\mathrm{AdS_2}$ case, and it is tempting to think that the Hamiltonian of the system is also related to some kind of unitary evolution of the microstates in the static patch. This may relate the finite dimension of de-Sitter quantum gravity Hilbert space to the fact that the mass of black holes in dS space cannot be infinite~\cite{Bousso_2000,Parikh_2005}. 

Apart from duality, in this article we also expolre other aspects of C-metric that may be helpful for future investigation of quantum gravity.

\begin{enumerate}

\item The thermodynamic variables and first law of the black hole. The biggest problem of presenting the first law of rotating C-metric lies on the definition of mass of the accelerating black hole. A well -known construction in~\cite{Dutta_2005} is the definition of Boost mass by extending the usual definition of ADM mass to asymptotic boosting case. Recently more studies are focused on considerations about the deficit angle~\cite{Appels_2016,rodríguez2021lawkerrtaubnutads,Bordo_2019,Liu_2022,gregory2020thermodynamicsblackholes,Anabal_n_2019}, while similar construction based merely on integrability with no cosmic string tension served as thermodynamic variables can be seen in~\cite{Astorino_2016}. These constructions still differ from each others in some ways. As pointed out in~\cite{Appels_2016}, some of them may face the problem of either being multi-boundary, which raises the problem of thermodynamic equilibrium. At first sight, a natural problem before making any hypothesis is the infinite area of the acceleration horizon, which require us to make up a regularization with physical rationality. In~\cite{PhysRevLett.75.3390,PhysRevD.51.4302} we have already seen some similar constructions, at least for the regularization of the area, and some similar treatment can be seen in~\cite{Ball_2021,Kim_2023}.

\item Reduction to 2D dilation-gravity model. It is believed that a universal property of extremal black hole is the AdS throat near horizon. For general spacetime accompanied with twist, of course, this description is too coarse: as we have mentioned before, we have NHEK for geometry for extremal Kerr BH and rotating Nariai for the extremal Kerr-dS or C-metric. Many interesting topics can be discussed once a spacetime is reduced to 2-dimension~\cite{Mertens_2023,Moitra_2019,Lam_2018,Yang_2019,saad2019latetimecorrelationfunctions,jafferis2023jtgravitymattergeneralized,harlow2019factorizationproblemjackiwteitelboimgravity,Brown_2019}, including a consistent quantum theory, information paradox and matrix integral duality. For this reason there is a necessity to conduct this reduction for more general case. We note that while statistics for extremal RN has been dealt with quite systematically~\cite{Iliesiu_2021}, there is still some problems for rotating case, such as, we can only preserve part of all metric fluctuation freedom. This leads to the lacking of a treatment on full quantum level. Still, it is beneficial to extend the results obtained before such as~\cite{Moitra_2019} to rotating C-metric.
\end{enumerate}

In this study, we provide a general and comprehensive description of the holography effect of the ``Nariai" flat C-metric solution. In Sec.\ref{sec:2}, we introduce the basic properties of rotating C-metric, which is aimed for mastering the macroscopic picture of the spacetime, which has very different properties compared to ordinary spacetime. Next part, Sec.\ref{sec:3}, would be a macroscopic and thermal investigation of the possible mass construction and first law of rotating C-metric based on Komar integral on the acceleration horizon through the background subtraction method and typical quasilocal construction, and although the formalism is quite different, the final result can precisely match some obtained before. This construction is inspired by some well-established treatment for the particle emission and entropy computation of this geometry~\cite{PhysRevLett.75.3390,PhysRevD.51.4302}. In Sec.\ref{sec:4}, in order to better interpret the thermal result microscopically, we also present the basic results related to holography properties of ``Nariai" flat C-metric to CFT, first at extremity and then beyond extremity for scalar field, so-called hidden symmetry. In Sec.\ref{sec:5}, for more general fluctuations in near-extremal rather than extremal solution, we reduce the rotating C-metric to 2 dimension to fit the form of Jackiw-Teitelboim gravity theory, and provide the bridge and foundation to the $n\mathrm{AdS_2}-n\mathrm{CFT_1}$, serving for possible further explorations in the future. 
 
\section{Basic properties of rotating C-metric}\label{sec:2}
The most general form of type-D electrovacuum solution family was firstly achieved by~\cite{Plebanski:1976gy}, and among them there exist one type of metric describing rotating and accelerating black hole with cosmological constant:
\begin{equation}\label{01}
\begin{aligned}
    \mathrm{d}s^2=\frac{1}{\Omega^2}\Big{[}-\frac{\Delta}{\rho^2}(\mathrm{d}t-a\sin^2\theta\Delta_{\phi}\mathrm{d}\phi)^2+\frac{\rho^2}{\Delta}\mathrm{d}r^2+\frac{\rho^2}{P}\mathrm{d}\theta^2
    +\frac{P\sin^2\theta}{\rho^2}[a\mathrm{d}t-(r^2+a^2)\Delta_{\phi}\mathrm{d}\phi]^2\Big{]}\,
\end{aligned}
\end{equation}
where
\begin{equation}\label{06}
    \Omega=1-\alpha r\cos\theta\,,\qquad P=1-2\alpha m\cos\theta+[\alpha^2(a^2+e^2)+\frac{1}{3}\Lambda a^2]\cos^2\theta\,,
\end{equation}
\begin{equation}\label{07}
    \rho^2=r^2+a^2\cos^2\theta\,,\qquad \Delta=(r^2-2mr+a^2+e^2)(1-\alpha^2r^2)-\frac{1}{3}\Lambda(a^2+r^2)r^2\,,
\end{equation}
with the vector potential for the gauged field given by
\begin{equation}
    A=-\frac{er}{r^2+a^2\cos^2\theta}(\mathrm{d}t+a\sin^2\theta\Delta_{\phi}\mathrm{d}\phi)\,,
\end{equation}
among three parameters, $a$, $e$ and $m$ are all interpreted as their usual meaning, i.e., the mass, the charge and the angular momentum-mass ratio. $\alpha$'s meaning is clear under massless and spinless case, which is the acceleration of the point particle at the origin of the coordinates. $\Delta_{\phi}$
is a constant to control the conical singularity at the north pole and south pole.

This metric has been intensively studied for many years~\cite{Ashtekar:1981ar,Griffiths:2006tk,Griffiths:2009dfa,Arenas_Henriquez_2022,arenashenriquez2023acceleratingblackholes21}, and here we briefly summarize some properties as well as physical interpretation of the solution. The original form of C-metric is supposed to describe two casually separated black holes which accelerate away due to the presence of cosmic strings, reflected by conical singularities. From Eq.(\ref{01}) it is obvious that the limit $\alpha=0$ responds to a common Kerr metric with two horizons, one Cauchy horizon and one event horizon. However, a rather small but nonzero $\alpha$ will cause tremendous alteration of the spacetime. 

First, the metric has a conformal factor $\Omega$ whose root corresponds to conformal infinity. Because the explicit meaning of all the coordinates, here we tacitly approve that the range of all coordinates are $0<r<\infty, 0<\theta<\pi, -\infty<t<\infty$. Then it's clear that when $0<\theta<\frac{\pi}{2}$ when $r$ approaches $\frac{1}{\alpha\cos\theta}$ we arrive at the conformal infinity $\mathcal{I}^{+}$, but when $\frac{\pi}{2}<\theta<\pi$ even when $r$ approaches to infinity we have no such boundary. Actually this time we can further extend the original coordinate beyond $r=\infty$ to $r=-\infty$ and then increases it until the point where $\Omega=0$. Another crucial difference from common solution is the existence of conical singularity. Take a $t,r=$ const spatial surface we get the induced metric
\begin{equation}\label{05}
    \mathrm{d}s^2=\frac{1}{\Omega^2}\Big{[}\frac{(r^2+a^2\cos^2\theta)}{P}\mathrm{d}\theta^2+\frac{P\sin^2\theta(r^2+a^2)^2-\Delta a^2\sin^4\theta}{r^2+a^2\cos^2\theta}\Delta_{\phi}^2\mathrm{d}\phi^2\Big{]}\,,
\end{equation}
and if we naively take $\Delta_{\phi}=1$ as $\theta\rightarrow 0$ and $\theta\rightarrow\pi$ we have deficit angle $\delta_{0}=2\pi[-\alpha^2(a^2+e^2)-\frac{1}{3}\Lambda a^2+2\alpha m]$ and $\delta_{\pi}=2\pi[-\alpha^2(a^2+e^2)-\frac{1}{3}\Lambda a^2-2\alpha m]$ respectively, which takes different value. Recall that there is also conical singularity of Kerr spacetime in Boyer-Lindquist coordinate, but as they take the same value when $\theta=0$ and $\theta=\pi$ we can easily move it out by redefining the period of the circulate coordinate $\phi$. Yet here things are more similar to Taub-Nut spacetime, when we can not move out two singularities at the same time unless we admit $t$ also have a certain period~\cite{Misner:1963fr}, which of course cause causality problems. Generally speaking, with a fixed $\Delta_{\phi}$, we must suppose the string with tension
\begin{equation}
    \mu_{\pm}=\frac{1}{4}\{1-[1+\alpha^2(a^2+e^2)+\frac{1}{3}\Lambda a^2 \pm 2m\alpha]\Delta_{\phi}\}\,.
\end{equation}
at the north pole ($\theta_{-}=0$) and south pole ($\theta_{+}=\pi$) respectively, and this time $\phi$ can always take its value in $(0,2\pi]$. 

\begin{figure}[htbp]
	\centering
\includegraphics[width=0.45\textwidth]{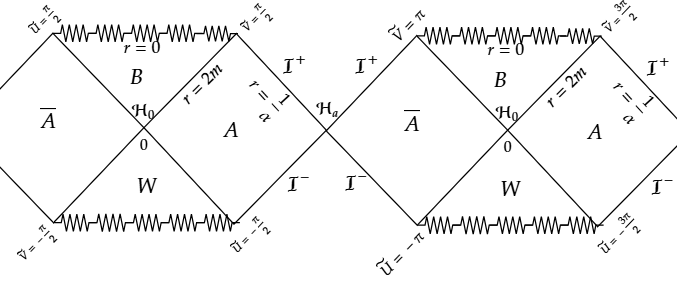}   
\includegraphics[width=0.43\textwidth]{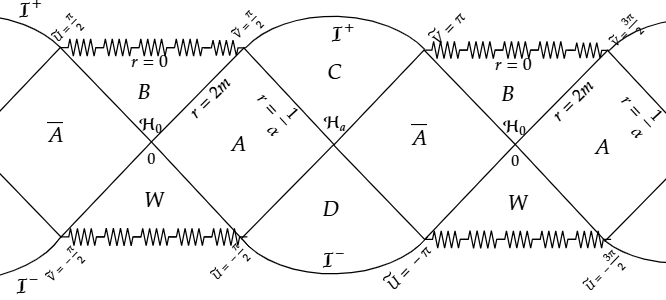}
\caption{The Penrose diagram for the global structure of C-metric in the case of $\theta=0$(left) and $0<\theta<\frac{\pi}{2}$ (right).}
\label{Fig.01}
\end{figure}

In order to clarify the global structure of the spacetime, we use the method introduced in~\cite{Griffiths:2006tk} to plot the conformal diagram of the C-metric. We first center on the ``flat" C metric with $\Lambda=0$, and for simplicity we also set $a=e=0$ and now the metric reduces to
\begin{equation}
    \mathrm{d}s^2=\frac{1}{\Omega^2}\Big{(}-Q\mathrm{d}t^2+\frac{\mathrm{d}r^2}{Q}+\frac{r^2\mathrm{d}\theta^2}{P}+Pr^2\sin^2\theta\mathrm{d}\phi^2\Big{)}\,,
\end{equation}
where
\begin{equation}
    Q=(1-\frac{2m}{r})(1-\alpha^2r^2)\,,\quad  P=1-2\alpha m\cos\theta.
\end{equation}
As usual we define $r_{*}=\int Q^{-1}\mathrm{d}r$ and $u=t-r_{*}, v=t+r_{*}$, and when in region $r\in(0,2m)$, we define $U=\exp({-\frac{\alpha u}{2\kappa_{0}}}), V=\exp({\frac{\alpha v}{2\kappa_{0}}})$, and now the metric turns to be 
\begin{equation}\label{02}
    \mathrm{d}s^2=-\Big{(}\frac{2\kappa_0}{\alpha}\Big{)}^2\frac{2m}{r(1-\alpha r\cos\theta)^2}(1-\alpha^2r^2)(1+\alpha r)^{-\frac{\kappa_{c}}{\kappa_{0}}}(1-\alpha r)^{-\frac{\kappa_{a}}{\kappa_{0}}}\mathrm{d}U\mathrm{d}V\,,
\end{equation}
in which
\begin{equation}
    \kappa_{0}=\frac{2\alpha m}{1-4\alpha^2m^2}\,,\qquad \kappa_{c}=\frac{1}{2(1+2\alpha m)}\,,\qquad \kappa_{a}=-\frac{1}{2(1-2\alpha m)}\,,
\end{equation}
and physical condition naturally requires that $\kappa_{c}>0, \kappa_{0}>0, \kappa_{a}<0$. We also have the relation 
\begin{equation}\label{03}
    UV=|1+\alpha r|^{\frac{\kappa_{c}}{\kappa_{0}}}|1-\alpha r|^{\frac{\kappa_{a}}{\kappa_{0}}}\Big{|}1-\frac{r}{2m}\Big{|}\,.
\end{equation}
If we further define conformal coordinates $\tan\Tilde{U}= U, \tan\Tilde{V}=V$, we see two special case: $r=0, UV=1, \Tilde{U}+\Tilde{V}=\frac{\pi}{2}$ (singularity) and $r=2m, UV=0, \Tilde{U}\Tilde{V}=0$ (event horizon). In this way we can plot  part $B$ of Fig.\ref{Fig.01} as well as its three boundaries. The rest work is quite similar. In the region $r\in(2m,\frac{1}{\alpha})$, we redefine $U$ and $V$ as $U=-\exp({-\frac{\alpha u}{2\kappa_{0}}}), V=\exp({\frac{\alpha v}{2\kappa_{0}}})$, (one can see more details in~\cite{Griffiths:2009dfa}), and we can preserve both the relation Eq.(\ref{02}) and Eq.(\ref{03}). As $\kappa_{a}<0$, $r=\frac{1}{\alpha}$ now corresponds to $UV=-\infty$, and either there is $\Tilde{V}=\frac{\pi}{2}, \Tilde{U}<0$ or $\Tilde{U}=-\frac{\pi}{2}, \Tilde{V}>0$. These are two accelerating horizons $\mathcal{H}_{a}$. In the region $\frac{1}{\alpha}<r<\infty$, we redefine 
$U=-\exp({-\frac{\alpha u}{2\kappa_{0}}}), V=-\exp({\frac{\alpha v}{2\kappa_{0}}})$. $r=\pm \infty$ correspond to $UV=0$ so is also a kind of horizon and when we pass it we need to either reverse the sign of $U$ or $V$. Similar extension can be made to cover every square horizontally or vertically until we encounter the conformal infinity.

\begin{figure}[htbp]
	\centering
\includegraphics[width=0.42\textwidth]{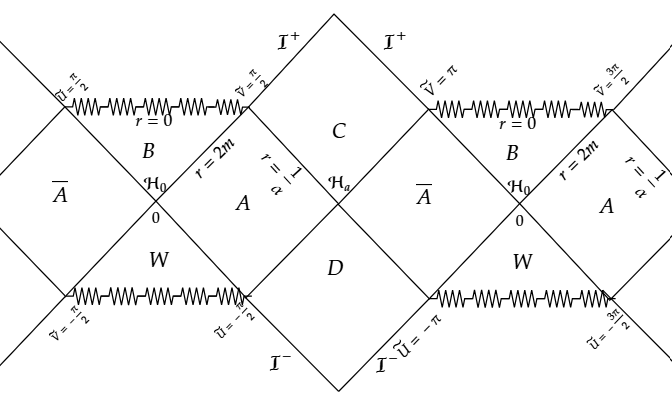}   
\includegraphics[width=0.27\textwidth]{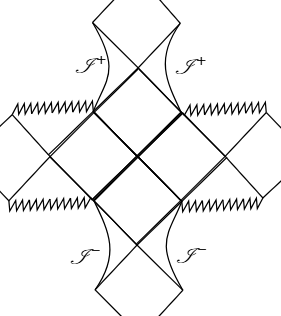}
\caption{The Penrose diagram for the global structure of C-metric in the case of $\theta=\frac{\pi}{2}$(left) and $\frac{\pi}{2}<\theta<\pi$ (right).}
\label{Fig.03}
\end{figure}

For the location of the conformal boundary, we are left with four possibilities: the first is when $\theta=0$, and the boundary is precisely where $r=\frac{1}{\alpha}$) as shown in the left panel of Fig.\ref{Fig.01}. The second case is when $\theta\in(0,\frac{\pi}{2})$, and the root of $\Omega$ satisfies a hypersurface which is spacelike:
\begin{equation}
\tan\tilde{U}\tan\tilde{V}=UV=\Big{(}1+\frac{1}{\cos\theta}\Big{)}^{\frac{\kappa_{c}}{\kappa_{0}}}(-1+\frac{1}{\cos\theta})^{\frac{\kappa_{a}}{\kappa_{0}}}(-1+\frac{1}{2m\alpha\cos\theta})\,,
\end{equation}
as shown in the right panel of Fig.\ref{Fig.01}. In the third case when $\theta=\frac{\pi}{2}$, we are left with a condition that the metric tensor component diverges when $r\rightarrow \infty $, so that's just where the boundary is, as shown in the left panel of Fig.\ref{Fig.03}. When $\theta\in(\frac{\pi}{2},\pi]$, the metric itself is finite even when $r$ approaches infinity. So a geodesic will pass it within finite proper time and the conformal boundary is even farther. Indeed, the location of this boundary is
\begin{equation}
    \tan\tilde{U}\tan\tilde{V}=UV=-\Big{(}-1-\frac{1}{\cos\theta}\Big{)}^{\frac{\kappa_{c}}{\kappa_{0}}}(1-\frac{1}{\cos\theta})^{\frac{\kappa_{a}}{\kappa_{0}}}(1-\frac{1}{2m\alpha\cos\theta})\,,
\end{equation}
which is a timelike boundary, as shown in the right panel of Fig.\ref{Fig.03}. Note now the spacetime can be extended even vertically.

The above discussion reveals that in the ``flat" C-metric case the acceleration horizon has the geometry of $\mathbb{R}\times\mathbb{R}^2=\mathbb{R}^3$ rather than usual horizon with $\mathbb{R}\times S^2$, for we have to cut out the north pole $\theta=0$ from the horizon, as it has extended to null infinity and does not belong to the spacetime. In conclusion, the nonzero $\alpha$ gives the whole structure some very different properties. The Penrose diagram is proved to be relied on the value of $\theta$, and as $\theta$ increasing from 0 to $\frac{\pi}{2}$ we can imagine the deformation of the boundary $\mathcal{I}^{+}$ of Fig.\ref{Fig.01} to that of Fig.\ref{Fig.03}. Moreover, we can even make more coordinate extension in the horizontal direction, especially when the rotation is also taken into account. 

\begin{figure}[htbp]
	\centering
\includegraphics[width=0.48\textwidth]{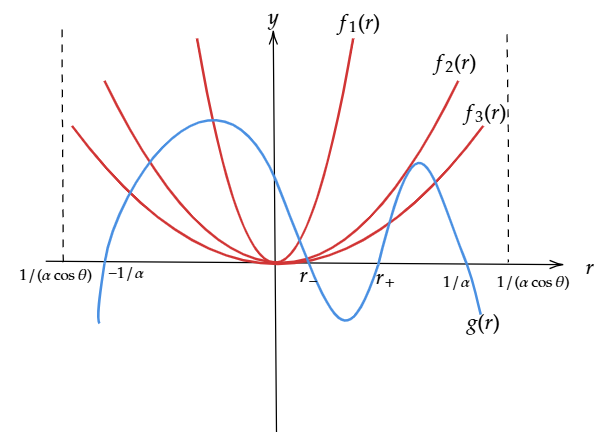}    \includegraphics[width=0.35\textwidth]{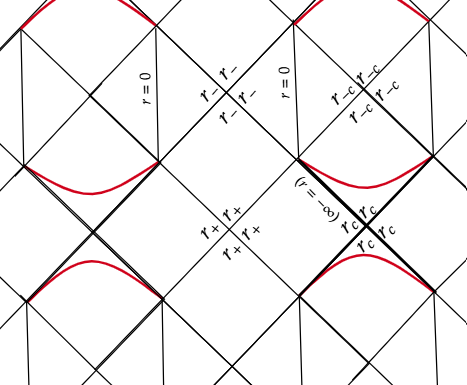}
\caption{Left:The three possible horizons under different parameters, where $f_{i}(r)=\frac{1}{3}\Lambda_i{} r^2(a^2+r^2)$,\, $g(r)=(r^2-2mr+a^2+e^2)(1-\alpha^2r^2)$, three different $\lambda$ correspond to one horizon, two degenerating horizons and three simple horizons in $r>0$ respectively. The possible conformal boundary is the dashed line. Right: The ``global" diagram of rotating C-metric with positive cosmological constant, where thick red lines correspond to conformal infinity and thick black lines are degenerate boundaries for two overlapping atlas.}
\label{Fig.05}
\end{figure}

We now present one more specific property of the C-metric. Using the form of the induced metric in Eq.(\ref{05}), we can calculate the topology of a given $t, r=$ const surface. The result is as expected:
\begin{equation}
\int_{S}R_{(2)}\epsilon_{ab}=8\pi[1+\alpha^2(a^2+e^2)]\,,
\end{equation}
where $S$ is any $t, r=$ const surface and $R_{(2)}$ is the scalar curvature of the surface, and the coordinate $\theta$ is supposed to vary from $0$ to $\pi$ on $S$. Now we can use Gauss-Bonnet theorem to attain
\begin{equation}
    \frac{1}{4\pi}\int_{S}R_{(2)}\epsilon_{ab}+\frac{\Delta \theta}{2\pi}=2\,,
\end{equation}
in which $\Delta\theta$ is the sum of the two deficit angle in the north pole and south pole. This confirms that $t, r=$ const surface is a 2-sphere even when $r$ approaches the acceleration horizon $\mathcal{H}_{a}$. This computation shows that every $r$-hypersurface can be interpreted as a deformed two-sphere $\tilde{S}^2$ except for the acceleration horizon; as shown in Fig.\ref{Fig.05}, it is exactly the intersection with the conformal boundary at $\theta=0$ that leads to this difference. 

Now the question arises that whether the ``flat" C-metric is in accordance to the definition of asymptotic flat spacetime. Although seemingly this metric has very different behavior when approaching no matter spatial infinity $i^{0}$ or null infinity $\mathcal{I}^{+}$, actually according to the study of~\cite{Ashtekar:1981ar,Dray:1982ex}, the most general C-metric (with both rotation and charge) satisfies the condition for an asymptotically empty and flat spacetime $(M,g_{ab})$ proposed in~\cite{Ashtekar:1978zz}:
\begin{enumerate}
    \item $(M,g_{ab})$ can be embedded in a larger spacetime $(\hat{M},\hat{g_{ab}})$, and existing a $C^{\infty}$ function $\Tilde{\Omega}$ on which satisfying $g_{ab}=\Tilde{\Omega}^2\hat{g_{ab}}$
    \item $\Tilde{\Omega}=0$ on $\partial \hat{M}$ while $\nabla \Tilde{\Omega}\neq 0$ on $\partial \hat{M}$.
     \item The manifold of orbits of the restriction of the vector field $n^{a}=\nabla^{a}\Tilde{\Omega}$ to  $\partial \hat{M}$ is diffeomorphic to $S^2$.
     \item $\Tilde{\Omega}^{-2}\hat{R_{ab}}$ has a smooth limit to $\partial \hat{M}$.
\end{enumerate}

After finishing the discussion mainly based on the asymptotically flat C-metric, we turn to the circumstance when the positive cosmological constant exists, and we expect the metric function $\Delta=(r^2-2mr+a^2+e^2)(1-\alpha^2r^2)-\frac{1}{3}\Lambda(a^2+r^2)r^2$ to have up to three positive roots. Now we bring the conformal boundary condition $r=1/(\alpha \cos\theta)$ into it and find that the boundary must be out of all the three horizons for any fixed $\theta$, as shown in Fig.\ref{Fig.05}, so at any extremal case (some horizons coincide), the transverse geometry will still be given by $\tilde{S}^2$. Also, the only possible degenerating will be between the outer (event) horizon and the cosmological (reduce to acceleration horizon in the ``flat" limit). In this general case we still need to extend the $r$ range to $[-\infty,0)$ just as in Kerr metric or asymptotically flat C-metric in order to keep consistency when passing through the 2-surface surrounded by the singular circle (this time the whole manifold will be a double-fold Riemann surface with fundamental group $\pi_1(M)=\mathbb{Z}_2$), and from Fig.\ref{Fig.05} there must be a negative root corresponding to another horizon behind $r=0$, and in both case we have to face the problem that the ``global" coordinate might also overlap; actually, as the surface gravity of the several horizons is not the same, we can only choose a coordinate covering two adjoint regions at one time, and the whole diagram is only a vivid illustration.

\section{ Deriving First law of rotating C-metic through background subtraction}\label{sec:3}

\subsection{Constructing thermodynamic variables}
Now we consider the general definition of Komar mass in a given spacetime. According to~\cite{PhysRev.113.934}, in an asymptotic flat and stationary spacetime (not necessarily static), one can always find a timelike killing vector $\xi^{a}$, and Komar mass can be defined as
\begin{equation}\label{komar}
    M_{S}=-\frac{1}{8\pi}\int_{S}\epsilon_{abcd}\nabla^{c}\xi^{d}\,,
\end{equation}
where here $S$ is a certain topological two-sphere. 

The crucial property of Komar mass is as follows:
\begin{equation}\label{04}
  M_{S}=M_{S_{\mathcal{H}}}-\frac{1}{4\pi}\int_{\Sigma}\epsilon_{abcd}R^{a}_{\; e}\xi^{e}\,,  
\end{equation}
where $\Sigma$ is the hypersurface whose boundaries are $S$ and a cross section of event horizon $S_{\mathcal{H}}$ respectively, and $R_{ab}$ is the Ricci tensor. This is natural considering the identities 
\begin{equation}
    \nabla_{a}k^a=0\,,\qquad \nabla_{a}\nabla^qk^b=-R^b_{\;a}k^a\,,
\end{equation}
and the first one permits the existence of Killing-Yano two form
\begin{equation}
    k^a=\nabla_{b}\omega^{ba}\,.
\end{equation}
Eq.(\ref{04}) argues that for a vacuum solution Komar mass will be a constant no matter how far we calculate it from the black hole horizon. Similar argument can make sense when we specifically consider a axial symmetric spacetime with two commutative killing vector $\xi^{a}, \phi^{a}$, and $\phi^{a}$ corresponds to a periodic coordinate, in which we can naturally extend this definition to Komar angular momentum:
\begin{equation}
    J_{S}=\frac{1}{16\pi}\int_{S}\epsilon_{abcd}\nabla^{c}\phi^{d}\,,
\end{equation}

Now before constructing the specific form of the mass, we review the method first introduced in~\cite{PhysRevD.51.4302}, i.e., by variable replacement
\begin{equation}
    (\alpha r)^{-1}=1+\epsilon (1-\chi)\,,\qquad \cos\theta=1-\epsilon\chi\,,
\end{equation}
to focus on the small region near the cosmic string in the north pole of the acceleration horizon, where $\epsilon$ is a infinitesimal and $\chi$ is considered as a coordinate function. Than on a $t=$ const hypersurface the metric can be rewritten in the form
\begin{equation}\label{asy}
    \mathrm{d}s^2=\frac{\Delta_{\phi}}{\epsilon}\Big{(}\frac{1}{\alpha^2}+a^2\Big{)}\Big{[}\frac{\mathrm{d}\chi^2}{2\chi(1-\chi)}+2\chi\mathrm{d}\phi^2+O(\epsilon)\Big{]}\,,
\end{equation}
where $\Delta_{\phi}=1/[1+2\alpha m+\alpha^2(a^2+e^2)]$. The spacetime is described by 4 parameters: $m, a, \alpha$ and $e$, yet in this limit we can abstractly owe the whole structure to three parameters: $\Delta_{\phi}, \epsilon$ and $u\equiv\frac{1}{\alpha^2}+a^2$. If we want to calculate true thermodynamical sums with a fixed background, we should always stay in a regularized frame  in which the asymptotic form of Eq.(\ref{asy}) is unchanged. Thus we force the following condition
\begin{equation}\label{19}
    \frac{\Delta_{\phi}u}{\epsilon}=\frac{\tilde{\Delta}_{\phi}\tilde{u}}{\tilde{\epsilon}}\,,
\end{equation}
which is generally different form some previous discussions of keeping the deficit angle in the north and south pole or equivalently, the tension of the string a constant. Now we aim to make a subtraction for the horizon area
\begin{equation}
\mathcal{A}_a^{(\epsilon)}=\int_{S}\sqrt{g_{\phi\phi}g_{\theta\theta}}\sin\theta\mathrm{d}\theta\mathrm{d}\phi=2\pi\Delta_{\phi}u(\frac{1}{\epsilon}-\frac{1}{2})\,,
\end{equation}
by taking
\begin{equation}
\Delta\mathcal{A}_a=\mathcal{A}_a^{(\epsilon)}-\tilde{A_a}^{(\epsilon)}=-\pi(\Delta_\phi u-\tilde{\Delta}_\phi\tilde{u})\,,
\end{equation}
and the question is how to eliminate the ambiguity in the second term. We have two constraints to make the subtraction: first is fixing the proper length of the boundary 
\begin{equation}
l_{-}=\int_{0}^{2\pi}\sqrt{g_{\phi\phi}}\mathrm{d}\phi=2\sqrt{2}\pi\sqrt{\frac{\Delta_\phi u}{\epsilon}}\sqrt{1-4\mu_{-}}+O(\epsilon)\,,
\end{equation}
which naturally leads to the condition $\mu_{-}=\tilde{\mu}_{-}$, where $\mu_{-}$ is the string tension at the north pole. Second is the Komar mass defined related to the subtraction can also have a finite sum. For simplicity we first consider the $a=e=0$ case, in which one may consider an integral very close to Eq.(\ref{komar}) but with a $\delta$ marking the beginning point of the integral:
\begin{equation}\label{inter}
    M=\frac{1}{2}\int_{\delta}^{\pi}\frac{\Delta_{\phi}\sin\theta\mathrm{d}\theta}{(1-\alpha r\cos\theta)^2}\Big{[}m(1-\alpha^2r^2)+r^2(1-\frac{2m}{r})\alpha\frac{\cos\theta-\alpha r}{1-\alpha r\cos\theta}\Big{]}\,,
\end{equation}
in which the $S$ in Eq.(\ref{komar}) has been chosen to be a $r=t=$ const surface so the $r$ in Eq.(\ref{inter}) is a fixed value. $\cos\delta=1-\epsilon$ is also a infinite small sum. For $r<\frac{1}{\alpha}$ it takes the value
\begin{equation}
    M_{bh}=\Delta_{\phi}[m+O(\epsilon)]\,,
\end{equation}
and as $\delta\rightarrow 0$ is precisely turns back to the expected value. But when $r=\frac{1}{\alpha}$, we have
\begin{equation}\label{reg}
    M_{a}^{(\epsilon)}=\Delta_{\phi}[\frac{1-2m\alpha}{2\alpha}(-\frac{1}{\epsilon}+\frac{1}{2})+O(\epsilon)]\,,
\end{equation}
note that this is a particular case for the 2-surface of acceleration horizon, which has the geometry of $\mathbb{R}^2$ rather than $S^2$. However, we can still take other 2-sphere which is as close as to the conformal boundary $1=r\alpha \cos\theta$ as possible and the integral still gives the value $M_{bh}=m\Delta_{\phi}$, and this is of course thanks to the vanishment of matter field. 

At the same time we can use the formula
$\kappa^2=-\frac{1}{2}\nabla^aK^b\nabla_{a}K_{b}\,,$
in which $\kappa$ is the surface gravity of a given killing horizon and $K^a$ is the corresponding killing normal vector of the horizon (here just $\xi^a=\partial_t$) to attain
\begin{equation}\label{sg}
\kappa_{bh}\equiv\kappa|_{r_{+}}=\frac{1-4\alpha^2m^2}{4m}\,,\qquad \kappa_{a}\equiv\kappa|_{r_{0}}=\alpha(1-2\alpha m)\,,
\end{equation}
in which $r_{0}=\frac{1}{\alpha}$ stands for the position of the acceleration horizon. And we also have the result
\begin{equation}
\mathcal{A}_{bh}\equiv\mathcal{A}|_{r_{+}}=\frac{4\pi\Delta_{\phi} m}{1-4\alpha^2m^2}\,,\qquad
\mathcal{A}_{a}^{(\epsilon)}=\frac{2\pi\Delta_{\phi}}{\alpha^2}(\frac{1}{\epsilon}-\frac{1}{2}+O(\epsilon))\,,
\end{equation}
and then we find the relation $M_a^{(\epsilon)}=-\frac{\kappa_a\mathcal{A}_a^{(\epsilon)}}{4\pi}$. Why there is a minus sign has no surprise if we recall the in the case of de-Sitter entropy calculation: there is also a minus sign~\cite{galante2023modavelecturenotessitter}. We can also have a formal derivation for the minus sign, in which actually we should change $\kappa_{a}$ in Eq.(\ref{sg}) to its opposite number. According to the original definition of the surface gravity, we have
\begin{equation}\label{kappa}
    Y_{a}\xi^a=-2\kappa Y_{a}\nabla^aa(\xi_{b}\xi^b)\,,
\end{equation}
in which $Y^a$ is a certain timelike and future-oriented vector located at the a certain horizon (say, one separating region $A$ from $C$ and one separating region $A$ from $B$ in Fig.\ref{Fig.01}), and then we notice that as $\xi^a$ is also future oriented on the second one while past-oriented on the first one. At the same time we have $\xi_{a}\xi^a<0$ in region $A$ and $\xi^a\xi_a>0$ in region $B$ and $C$,  so $Y_{a}\nabla^a(\xi_{b}\xi^b)>0$ and we have $\kappa>0$ on the first one while $\kappa<0$ for the second one. In this sense we will absorb the minus sign into $\kappa_a$ in the following thermodynamic identities, or equivalently, we are computing the effective Komar mass between the acceleration horizon and conformal infinity. This choice perhaps has no much physical meaning but is essential for the construction of the first law.

Now with the interpretation of Eq.(\ref{19}) we may define the subtracted Komar mass by requiring that the Smarr relation always makes sense:
\begin{equation}
    \Delta M_a\equiv \frac{1}{4\pi}(\kappa_a\mathcal{A}_a^{(\epsilon)}- \tilde{\kappa}_a\tilde{\mathcal{A}}^{(\epsilon)}_a)\,, 
\end{equation}
then we require $\kappa_a=\tilde{\kappa}_a$ for $\Delta M_a$ to be regular. Then we get $\Delta \phi/\alpha =\tilde{\Delta}_{\phi}/\tilde{\alpha}$ and if at this time we further require $\mu_{+}=\mu_{+}$ (This is equivalent to say $\alpha=\tilde{\alpha}$), we naturally attain $\Delta_{\phi}=\tilde{\Delta}_{\phi}$ and $\Delta M_a=\Delta \mathcal{A}_a=0$. Thus we get similar results in some previous papers, in which ``boost mass" corresponding to the acceleration horizon is always 0. This trivial result, however, is based on fixing the conical deficit at both poles, and provide us with the intuition that the final first law expression must contain variation of these two variables. 

For more general rotating and charged C-metric in Eq.(\ref{01}), we can still take the definition and requirement above, but we further impose $\Phi_a=\tilde{\Phi}_a$ as well.  We first compute the surface gravity at the acceleration horizon to be
\begin{equation}
    \kappa_a=-\frac{1-2\alpha m+\alpha^2(a^2+e^2)}{\alpha u}\,,
\end{equation}
and then $\kappa_a=\tilde{\kappa}_a, \; \Phi_a=\tilde{\Phi}_a$ and $\mu_{-}=\tilde{\mu}_{-}$ lead to 
\begin{equation}
    q\equiv \alpha(1+s^2\alpha^2u)-2\frac{m}{u}=\tilde{\alpha}(1+s^2\tilde{\alpha}^2\tilde{u})-2 \frac{\tilde{m}}{\tilde{u}}\,, \qquad s\equiv \frac{e}{u\alpha}=\frac{\tilde{e}}{\tilde{u}\tilde
    \alpha}\,, \qquad f=\alpha u\Delta_{\phi}=\tilde{\alpha}\tilde{u}\tilde{\Delta}_{\phi}\,,
\end{equation}
and then there is $\kappa_a=\tilde{\kappa}_a=-q$, so totally we have three independent constraints $\{q,s,f\}$ for five parameters $\{m,a,e,\alpha,\Delta_{\phi}\}$.

If we compute the Komar integral as in Eq.(\ref{komar}) at the acceleration horizon,  we have the explicit form of the background subtracted mass as
\begin{equation}
    \Delta M_a=2(\frac{m\Delta_\phi a^2}{u}-\frac{\tilde{m}\tilde{\Delta}_\phi \tilde{a}^2}{\tilde{u}})+\frac{qf}{4}(\frac{1}{\alpha}-\frac{1}{\tilde{\alpha}})\,,
\end{equation}
which satisfies the Smarr relation
\begin{equation}\label{lll}
    \Delta M_a=2(\Omega_aJ_a-\tilde{\Omega}_a\tilde{J}_a)+\frac{1}{4\pi}(\kappa_a\mathcal{A}_a^{(\epsilon)}- \tilde{\kappa}_a\tilde{\mathcal{A}}^{(\epsilon)}_a)\,.
\end{equation}

There is still two degrees of freedom in confirming the background spacetime and  a natural choice would be $\tilde{a}=0, \alpha=\tilde{\alpha}$.
Then the subtraction gives the difference of the mass between rotating and non-rotating BHs with the same acceleration.
Note that under this subtraction scheme the second term of the RHS of Eq.(\ref{lll}) is always 0 and so we can merely leave the first term when discussing saddle-point perturbation restricted on the orbit of this family of solutions with varying parameters. Generally  we will have the following relation
\begin{equation}
    \tilde{\alpha}=\alpha\,, \quad \tilde{a}=0, \quad \tilde{e}=\frac{e}{1+\alpha^2a^2}\,, \quad \tilde{\Delta}_{\phi}=(1+\alpha^2a^2)\Delta_{\phi}\,,\quad \tilde{m}=\frac{1}{1+\alpha^2 a^2}[m-\frac{e^2a^2\alpha^3}{2(1+a^2\alpha^2)}]\,,
\end{equation}
and the explicit expression for the subtracted mass is
\begin{equation}
    \Delta M_a=\frac{2\Delta_{\phi}\alpha^2a^2m}{1+a^2\alpha^2}=2\Omega_aJ_a\,.
\end{equation}

On the other hand, for the Komar integral of $M$ at the conformal infinity there is
\begin{equation}
    M_{bh}=m\Delta_\phi\,,
\end{equation}
while
\begin{equation}
\mathcal{A}_{bh}=4\pi\Delta_{\phi}\frac{r_{+}^2+a^2}{1-(\alpha r_{+})^2}\,, \quad \kappa_{bh}=\frac{[1-(\alpha r_{+})^2][mr_{+}-(a^2+e^2)]}{r_{+}(r_{+}^2+a^2)}\,,
\end{equation}
 and
\begin{equation}
    J_{a}=\Delta_{\phi}^2ma\,,\qquad \Omega_{bh}=\frac{\Delta_{\phi}^{-1}a}{a^2+r_{+}^2}\,,\qquad \Phi_{bh}=\frac{er_{+}}{a^2+r_{+}^2}\,,
\end{equation}
 so the Smarr relation
\begin{equation}\label{smarr}
{M}_{bh}-2\Omega_{bh}J_{a}-\Phi_{bh}\mathcal{Q}=\frac{\kappa_{bh}\mathcal{A}_{bh}}{4\pi}\,,
\end{equation}
is also satisfied. Now we define $M$ to be the mass ``between the acceleration horizon and conformal infinity", i.e., $\Delta M\equiv {M}_{bh}-\Delta M_{a}$, which satisfies the total Smarr relation
\begin{equation}
    \Delta M=2(\Omega_{bh}-\Omega_a)J_a+\Phi_{bh}\mathcal{Q}+\frac{\kappa_{bh}A_{bh}}{4\pi}\,.
\end{equation}

However, the definition above is still not enough to get an integrable first law, unless we make a further gauge $\mathrm{d}t\rightarrow \mathrm{d}t/\gamma$, but with no change in Killing vector to compute thermodynamic variables, then we have to make the replacement
\begin{equation}
    \Delta M \rightarrow \frac{\Delta M}{\gamma}\,, \qquad \Omega \rightarrow \frac{\Omega}{\gamma}\,, \qquad \Phi \rightarrow \frac{\Phi}{\gamma}\,, \qquad \kappa \rightarrow \frac{\kappa}{\gamma}\,,
\end{equation}
these are to keep the original defined relation invariant. If we take the arbitrary 
factor $\gamma$ to be
\begin{equation}
    \gamma=\frac{\sqrt{(1-a^2\alpha^2)[1+\alpha^2(a^2+e^2)]}}{1+a^2\alpha^2}\,,
\end{equation}
then we are able to obtain a form of first law as
\begin{equation}\label{first law}
    \delta \Delta M= \frac{\kappa_{bh}}{8\pi}\delta \mathcal{A}_{bh}+ (\Omega_{bh}-\Omega_{a})\delta J_{a}+\Phi_{bh}\delta \mathcal{Q}-\lambda_{-}\delta \mu_{-}-\lambda_{+}\delta\mu_{+}\,,
\end{equation}
which precisely coincides with the form obtained in~\cite{Anabal_n_2019}. As mentioned there, the reason for parameterizing $t/\gamma$ is unclear, at least for asymptotically flat case. Here $\lambda_{\pm}$ is the thermodynamic lengths corresponding to the deficit angle at two poles respectively~\cite{Ball_2021}. Their value can be explicitly written as
\begin{equation}
    \lambda_{\pm}=\frac{r_{+}}{\gamma(1\mp\alpha r_{+})}-\frac{M}{\Delta_{\phi}(1+\alpha^2(a^2+e^2))}\pm \frac{a^2\alpha}{\gamma(1+\alpha^2a^2)}\,.
\end{equation}
Here it is also worthy to mention that in~\cite{Anabal_n_2019}, the mass formula is defined through the transformation $\xi^a\rightarrow \xi^a+\Omega_a \phi^a$ when computing the Komar integral for at infinity. Though we generally construct the mass term in a relatively different way, their definition is better for us to derive the first law in with more general condition as we will show in the next subsection.
                                        
\subsection{Deriving the first law using covariant phase space formalism}
\begin{figure}[htbp]
	\centering
\includegraphics[width=0.5\textwidth]{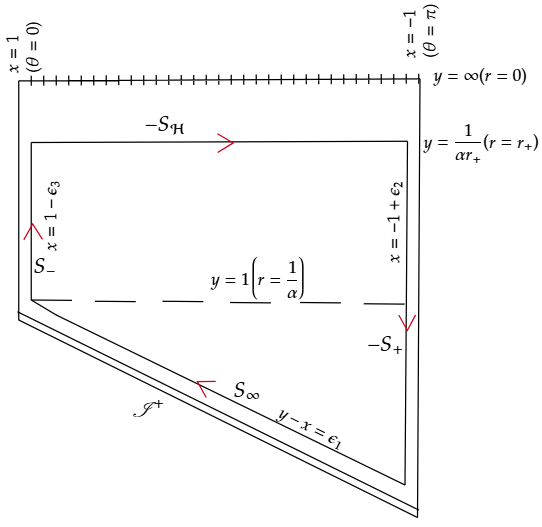}   
\caption{The sketch map of the Cauchy slice and its subset $\Sigma$, in which the whole loop is $\partial \Sigma=(-S_{\mathcal{H}}) \cup (-S_{+}) \cup S_{-}\cup S_{\infty}$, and $\phi$ coordinate is suppressed.}
\label{Fig.04}
\end{figure}

We now derive the first law formally through the covariant phase space formalism~\cite{Wald_1993,Iyer_1994}. This formalism is based on covariant analyses, but in practical we still take some gauge to simplify the discussion, as although the abstract form of the law is gauge invariant, the explicit form is not. Now make the coordinate transformation $y=\frac{1}{\alpha r}, x=\cos\theta$, the global structure of the spacetime is shown in Fig.\ref{Fig.04}.

In~\cite{Wald_1993}, it is proved that for a spacetime with two symmetries, the symplectic form constructed from the linear combination of these two Killing vectors $\chi^a$ vanishes
\begin{equation}
0=\omega[\delta,\delta_{\chi}]=\delta\Theta[\psi,\mathcal{L}_{\chi}\psi]-\mathcal{L}_{\chi}\Theta[\psi,\delta\psi]\,,
\end{equation}
there $\psi$ is a dynamical field which could be metric $g_{ab}$ or gauge field $A^a$ in Einstein-maxwell theory. This gives rise to a conserved charge through
\begin{equation}\label{waldf}
0=\int_{\Sigma}\omega[\delta,\delta_{\chi}]=\int_{\partial\Sigma}\delta Q[\chi]-\chi\cdot \Theta[\delta]\,,
\end{equation}
where $\Sigma$ can be spacelike subset of a Cauchy slice and $\partial\Sigma$ is its boundary. One can divide it into four parts $\partial\Sigma=(-S_{\mathcal{H}}) \cup (-S_{+}) \cup S_{-}\cup S_{\infty}$, in which the $S_{\mathcal{H}}$ is a section of the event horizon, $S_{\pm}$ are thin tubes around north (south pole) and $S_{\infty}$ is a section of conformal infinity (see Fig.\ref{Fig.04} for detail). In Einstein-Maxwell theory, $Q[\chi]$ and $\Theta[\chi]$ are functionals of the Killing vector $\chi^a$, which are the sum of two terms coming from contributions of gravity and electromagnetic field respectively. The explicit forms are
\begin{equation}
    Q[\chi]=Q_{\mathrm{grav}}[\chi]+Q_{\mathrm{em}}[\chi]\,,\qquad \Theta[\delta]=\Theta_{\mathrm{grav}}[\delta]+\Theta_{\mathrm{em}}[\delta]\,,
\end{equation}
in which 
\begin{equation}
    Q_{\mathrm{grav},ab}[\chi]=-\frac{1}{16\pi}\epsilon_{abcd}\nabla^c\chi^d\,,
\end{equation}
\begin{equation}
    \Theta_{\mathrm{grav},abc}[\delta]=\frac{1}{16\pi}\epsilon_{abcd}g^{de}g^{fh}(\nabla_f\delta g_{eh}-\nabla_{e}\delta g_{fh})\,,
\end{equation}
and 
\begin{equation}
    Q_{\mathrm{EM},ab}[\chi]=-\frac{1}{32\pi}\epsilon_{abcd}F^{cd}\chi^fA_f\,,
\end{equation}
\begin{equation}
    \Theta_{\mathrm{EM},abc}[\delta]=\frac{1}{16\pi}\epsilon_{abcd}F^{fd}\delta A_{f}\,.
\end{equation}
Note here the gravity part of $Q_{ab}$ is nothing but Komar integral.

Now we can explicitly compute every term in Eq.(\ref{waldf}) and reveal their relation to thermodynamic variables. We make the decomposition $\chi^a=\chi_1^a+\chi_2^a\equiv(\xi^a+\Omega_a\phi^a)+(\Omega_{bh}-\Omega_a)\phi^a$. Then one defines the Hamiltonian ``energy" (in asymptotic Minkowski spacetime ADM integral) as 
\begin{equation}
    \slashed \delta H[\chi_1]=\int_{S_{\infty}}\delta Q_{\mathrm{grav}}[\chi_1]-\chi_1\cdot \Theta_{\mathrm{grav}}[\delta]\,,
\end{equation}
and here $\slashed\delta$ means it is perhaps not really integrable. On the other hand,
\begin{equation}
     \delta H[\chi_2]=\int_{S_{\infty}}\delta Q_{\mathrm{grav}}[\chi_2]-\chi_2\cdot \Theta_{\mathrm{grav}}[\delta]=\int_{S_{\infty}}\delta Q_{\mathrm{grav}}[\chi_2]\,,
\end{equation}
gives just the Komar integral with respect to $\phi^a$. The possibly nonintegrable part vanishes as we choose $S_{\infty}$ to be $\{y-x=\epsilon,0\leq \phi\leq 2\pi\}$
and thus $\phi^a$ is tangential to it. Now under the gauge $\Omega_{bh}-\Omega_{a}$ is fixed, we claim that
\begin{equation}
    \delta H[\chi_1]=\delta \Delta M, \qquad \delta H[\chi_2]=-(\Omega_{bh}-\Omega_a)\delta J_a\,,
\end{equation}
Regarding that here no energy flux is expected to leak into null infinity (such as in Bondi-Sach), the integrability of the pre-symplectic form $\slashed \delta H[\chi_1]$ is reasonable. Moreover, asymptotic flatness guarantees that ADM mass equals Komar mass. One can verify this through direct computation. Thus 
\begin{equation}
    \slashed \delta H[\chi]= \int_{S_{\infty}}\delta Q_{\mathrm{grav}}[\chi]-\chi\cdot \Theta_{\mathrm{grav}}[\delta]=\delta \Delta M-(\Omega_{bh}-\Omega_a)\delta J_a\,.
\end{equation}

On the other hand, because $\chi^a=\xi^a+\Omega_{bh}\phi^a$ is the Killing vector of Killing horizon and thus vanishes on it, following the proof in~\cite{Wald_1993}, we have
\begin{equation}
\int_{S_{\mathcal{H}}}\delta Q_{\mathrm{grav}}[\chi] -\chi\cdot \Theta[\delta]=\int_{S_{\mathcal{H}}}\delta Q_{\mathrm{grav}}[\chi]= \frac{\kappa_{bh}}{8\pi}\delta \mathcal{A}_{bh}\,,
\end{equation}
given that $\kappa_{bh}$ is fixed at the horizon.

The contributions from the integral on the two tubes turn out to be
\begin{equation}
    \int_{S_{\pm}} \delta Q_{\mathrm{grav}}[\chi]-\chi\cdot \Theta_{\mathrm{grav}}[\delta]=-\int_{S_{\pm}}\chi\cdot\Theta[\delta]=-\int_{S_{\pm}}\xi\cdot \Theta[\delta]=\mp \lambda_{\pm}\delta \mu_{\pm}\,.
\end{equation}
The first equals sign makes sense because while $\sqrt{-g}=0$ in two tubes ($\{r,0\leq \phi\leq 2\pi\}$), $g^{\theta\theta}$ and $\Gamma^{t(\phi)}_{\theta\mu}$ are all regular and thus $\nabla^{\theta}\chi^t$ is regular and thus the integral vanishes. The second makes sense because $\phi^a$ is tangential to the tubes. 

For the gravitational part, it's also meaningful to mention the work in~\cite{Kim_2023}, in which a renormalized symplectic potential as well as Hamiltonian charge is defined in slowly rotating black hole in asymptotic $\mathrm{AdS}$ spacetime. However for flat accelerating black hole the contribution of counter term to symplectic potential is found to vanish. This is also supported by the computation of ADM mass at conformal infinity to converge.

We now turn to the EM part, in which there is~\cite{Kim_2023}
\begin{equation}
    \delta Q_{\mathrm{EM}}-\chi\cdot \Theta_{\mathrm{EM}}=-\frac{1}{4\pi}(\chi\cdot A) \delta (*F)\,,
\end{equation}
and thus without imposing the potential at the horizon we are able to have
\begin{equation}
    \int_{S_{\mathcal{H}}}  \delta Q_{\mathrm{EM}}-\chi\cdot \Theta_{\mathrm{EM}}= \Phi_{bh}\delta \mathcal{Q}\,,
\end{equation}
and at conformal infinity the integral of EM part is obviously zero, as $\lim _{\mathcal{I^+}}\Phi=0$. Also, as there is no electric charge contained inside the tube, the contribution of the integral on $(-S_{+})\cup S_{-}$ vanishes. 

When we combine all the results obtained before and bring them into Eq.(\ref{waldf}), we get exactly the formula in Eq.(\ref{first law}). In the proof we still need to take the usual gauge condition for covariant phase space, in which $\Omega_{bh}-\Omega_{a}$, as well as $\kappa_{bh}$ is fixed. For this reason the main result in this subsection is actually different from proving through the total variation in parameter space, as here all variables can have geometric interpretation and the first law can make sense for any type of perturbation (such as time dependent ``hair") but obeying our gauge condition. 

\section{Holography dual of rotating C-metric in Nariai limit}\label{sec:4}
\subsection{Extremal duality}
After obtaining the thermodynamic law for flat C-metric, we provide the possible microscopic explanation to it through the formalism similar to Kerr/CFT correspondence. We first consider the specific form of the metric in the ``Nariai" flat limit~\cite{Anninos_2010,Booth_1999}.  From the expression in Eq.(\ref{01}), we see generally there should be three horizons, at the location $r=r_{-}$, $r=r_{+}$ and $r=\frac{1}{\alpha}$, in which $r_{-}$ and $r_{+}$ satisfy the relation $r_{+}r_{-}=a^2+e^2$, $r_{+}+r_{-}=2m$. Now we take the limit in which $r_{+}$ approaches $\frac{1}{\alpha}$, rather than the usual treatment of $r_{-}\rightarrow r_{+}$~\cite{Astorino_2016}, so we attain a ``Nariai" flat C metric, which has a different global structure from that has been comprehensively discussed in literature before. In this limit the definition of those metric functions in Eq.(\ref{06}) and Eq.(\ref{07}) reads as
\begin{equation}
    \Omega=1-\alpha r\cos\theta\,,\qquad P=(1-\cos\theta)(1-\alpha r_{-}\cos\theta)\,,
\end{equation}
\begin{equation}
    \rho^2=r^2+a^2\cos^2\theta\,,\qquad \Delta=-\frac{1}{\alpha}(r-r_{-})(1-\alpha r)^2(1+\alpha r)\,,
\end{equation}
 Now we focus on the near-horizon limit of this solution. As usual we introduce the dimensionless coordinates as in~\cite{PhysRevD.80.124008}:
\begin{equation}\label{trans}
     \hat{r}=\frac{\alpha r-1}{\lambda}\,,\qquad \hat{t}= \frac{\lambda t}{b}\,,\qquad \hat{\phi}=\phi-\tilde{b}t\,,
\end{equation}
in which 
\begin{equation}
    b=\frac{1+\alpha^2a^2}{2(1-\alpha r_{-})\alpha}\,,\qquad
    \tilde{b}=\frac{a\alpha^2\Delta_{\phi}^{-1}}{1+\alpha^2a^2}\,,
\end{equation}
and for simplicity from now on we omit the hat of the new coordinates. The Nariai-type metric is now 
\begin{equation}\label{11}
    \mathrm{d}s^2=\Gamma(\theta)\Big{[}r^2\mathrm{d}t^2-\frac{\mathrm{d}r^2}{r^2}+\gamma(\theta)\mathrm{d}\theta^2+\beta(\theta)(\mathrm{d}\phi+\zeta r\mathrm{d}t)^2\Big{]}
\end{equation}
where
\begin{equation}
    \Gamma(\theta)=\frac{(1+\alpha^2a^2\cos^2\theta)}{2\alpha^2(1-\cos\theta)^2(1-\alpha r_{-})}\,,\qquad
    \gamma(\theta)=\frac{2(1-\alpha r_{-})}{(1-\cos\theta)(1-\alpha r_{-}\cos\theta)}
\end{equation}
and
\begin{equation}
    \beta(\theta)=\frac{2(1-\alpha r_{-})(1-\cos\theta)(1-\alpha r_{-}\cos\theta)(1+a^2\alpha^2)^2\sin^2\theta}{(1+\alpha^2a^2\cos^2\theta)^2}\Delta_\phi^2
\end{equation}
\begin{equation}
    \zeta=\frac{a\alpha}{(1+a^2\alpha^2)(1-\alpha r_{-})\Delta_\phi}\,.
\end{equation}
Eq.(\ref{11}) is precisely the type of Nariai limit~\cite{galante2023modavelecturenotessitter}, except for one difference that the range of $\theta$ is now $(0,\pi]$ rather than $[0,\pi]$ so the whole geometry of constant $t,r$ slice is now $\mathbb{R}^2$ rather than $S^2$(usual Nariai geometry) or $\tilde{S}^2$ (Nariai C metric where there is conical singularity at one or two poles). Clearly this is because when we are in ``Nariai" flat limit the north pole (where $\Gamma(\theta)$ and other metric functions diverges) has been cut out from the whole manifold. Meanwhile, the gauge field reads as (after a gauge transformation to move out a infinite constant)
\begin{equation}
    A=-\frac{e}{1+a^2\alpha^2\cos^2\theta}\Big{[}\frac{1-a^2\alpha^2\cos^2\theta}{2(1-\alpha r_{-})}r\mathrm{d}t+a\alpha\sin^2\theta\Delta_{\phi}\mathrm{d}\phi\Big{]}\,,
\end{equation}

The isometry group of this geometry is generated by
\begin{equation}
    \xi_{-1}=\Big{(}\frac{\partial}{\partial t}\Big{)}^a\,,\qquad \xi_{0}=t\Big{(}\frac{\partial}{\partial t}\Big{)}^a-r\Big{(}\frac{\partial}{\partial r}\Big{)}^a\,,
\end{equation}
\begin{equation}
    \xi_{1}=\Big{(}\frac{1}{2r^2}+\frac{t^2}{2}\Big{)}\Big{(}\frac{\partial}{\partial t}\Big{)}^a-tr\Big{(}\frac{\partial}{\partial r}\Big{)}^a-\frac{\zeta}{r}\Big{(}\frac{\partial}{\partial \phi}\Big{)}^a\,,\qquad L_{0}=\Big{(}\frac{\partial}{\partial \phi}\Big{)}^a\,,
\end{equation}
which satisfies the $SL(2,R)\times U(1)$ algebra:
\begin{equation}
    [\xi_{0},\xi_{\pm 1}]=\pm\xi_{\pm 1}\,,\qquad [\xi_{-1},\xi_{+1}]=\xi_{0}\,.
\end{equation}
in which $\xi_{\pm1,0}$ serve as the $SL(2,R)$ generators while $L_{0}$ serves as the $U(1)$ generator. This is precisely the whole generators of the symmetric group of warped $\mathrm{CFT}_{2}$, which is the same group we discuss in usual Kerr-CFT and rotating Nariai-CFT. This has no coincidence as in the above expression it is clear that only difference between the ``Nariai" flat and dS Nariai is the compactness on $\theta=0$ side which has no influence on the local Killing equation. On the other hand, asymptotically dS Nariai geometry enjoys the same isometry group with NHEK, which is asymptotically AdS. Because the warped $\mathrm{dS}_3$ geometry is fulfilled at every $\theta$-slice which can be viewed as a radius-deforming fiberation of $S_1$ over $\mathrm{dS_2}$, the whole isometry group breaks from $SL(2,R)_L\times SL(2,R)_R$ (that of pure $\mathrm{dS}_3$) to exactly $SL(2,R)\times U(1)$. The identification of $\phi$ and $\phi+2\pi$ plays the role of taking finite temperature when discussing the dual conformal field theory~\cite{Afshar_2020,Aggarwal_2023}. The next step is to confirm the asymptotic symmetry group (ASG) of the geometry. ASG is defined as quotient group:
\begin{equation}
    \mathrm{ASG}=\frac{\mathrm{All\; allowed\; diffeomorphisms}}{\mathrm{Trivial\; diffeomorphisms}}\,,
\end{equation}
where ``Allowed" restricts the generator of the diffeomorphism to preserve certain asymptotic condition of the spacetime, while ``trivial'' means that the generator of the transformation vanishes after we have implemented the constraints and reduced it to a boundary integral. We can also understand it by the following symplectic structure:
\begin{equation}\label{defi}
    \left\{Q_{\xi},\Phi\right\}=\mathcal{L}_{\xi}\Phi\,,
\end{equation}
in which $\left\{\,,\,\right\}$ denotes the Dirac brackets, when there exists constraints in the phase space of the system. Then in General Relativity the calculation of $Q_{\xi}$ is typically only effective as the boundary terms, and when the boundary integration gives zero value it will be a trivial diffeomorphism. Here as first pointed out in~\cite{PhysRevD.80.124008}, to keep the asymptotic structure of Nariai geometry, the perturbation from the original metric should be
\begin{equation}\label{b-c}
    h_{\mu\nu}\sim \mathcal{O}\begin{pmatrix}r^2&1&1/r&1/r^2\\&1&1/r&1/r\\&&1/r&1/r^2\\&&&1/r^3\end{pmatrix}\,,
\end{equation}
and thus the most general generator of the diffeomorphism to preserve the asymptotic condition is
\begin{equation}
\xi_{\epsilon}=\epsilon(\phi)\Big{(}\frac{\partial}{\partial \phi}\Big{)}^a-r\epsilon'(\phi)\Big{(}\frac{\partial}{\partial r}\Big{)}^a\,,
\end{equation}
and we can express it in the basis $\xi_{n}=\xi(-\exp^{-in\phi})$ which satisfies the Virasaro algebra
\begin{equation}\label{com}
    i[\xi_{m},\xi_{n}]=(m-n)\xi_{m+n}\,,
\end{equation}
and the next question is to confirm the central charge of this CFT. The Virasaro algebra here only contains a $U(1)$ but not $SL(2,R)$ subgroup, which suggests that the CFT state dual to the Kerr vacuum is not $SL(2,R)$ invariant. At the quantum level, Eq.(\ref{com}) can be applied central extension to satisfy the most general form of Virasaro algebra
\begin{equation}
    [L_m,L_n]=(m-n)L_{m+n}+\frac{c}{12}m(m^2-1)\delta_{m+n,0}\,,
\end{equation}
and that means we can compute central charge through
\begin{equation}
c_L=12i\int_{\partial\Sigma}k_{\xi_{m}}[\mathcal{L}_{\xi_{-m}}g,g]\,,
\end{equation}
where
\begin{equation}
    k_{\xi}[h,g]=(\xi^{b}\nabla^ah+\xi_{c}\nabla^bh^{ca}+\xi^a\nabla_{c}h^{cb}+\frac{1}{2}h\nabla^a\xi^b+\frac{1}{2}h^{ac}\nabla_{c}\xi^b+\frac{1}{2}h^{bc}\nabla^a\xi_c)dx_a\wedge dx_b\,.
\end{equation}
is a sum defined in classical gravity theory, which satisfies the relation
\begin{equation}\label{central ex}
\{Q_{\xi_m},Q_{\xi_n}\}_{\mathrm{D.B.}}=Q_{[\xi_m,\xi_n]}+\frac{1}{8\pi}\int_{\partial \Sigma} k_{\xi_m}[\mathcal{L}_{\xi_n}g,g]\,,  
\end{equation}
and here all covariant derivatives correspond to $g_{ab}$~\cite{Barnich_2002}. It is well-known that the idea that canonical realization like $Q[\xi]$ of asymptotic symmetry can induce a central extension in its Lie Algebra just as in Eq.(\ref{central ex}) was first proposed by Brown and Henneaux for $\mathrm{AdS}_3$ gravity~\cite{Brown:1986nw}. For those diffeomorphism that are not exact symmetry of background geometry or pure gauge (related to trivial action), the central extension is always nontrivial. The reason we encounter only a chiral half of a CFT ultimately derives from the fact that at extremity the rotational velocity of the Kerr-type horizon becomes the speed of light. This forces all physical excitations (such as the edge of the accretion disc) to spin around chirally with the black hole. Specifically, in spacetime of the form Eq.(\ref{11}), we can get the central charge possibly dual to a chiral CFT as
\begin{equation}
c_L=3\zeta\int_{0}^{\pi}\Gamma(\theta)\sqrt{\beta(\theta)\gamma(\theta)}\mathrm{d}\theta\,,
\end{equation}
Now this central charge term is again a positive infinite value, which accords to the requirement of near-extremal limit of the Cardy formula for entropy calculating. In the regulation scheme of Eq.(\ref{reg}), its regular term reads as
\begin{equation}
    c_L^{(\epsilon)}=-\frac{3a}{\alpha(1-\alpha r_{-})}\,.
\end{equation}

We now turn to the temperature. Although in non-extremal case there exists two different temperatures, in extremal limit there is no such problem~\cite{Booth_1999}. Thus we can choose the temperature on the event horizon and write in the expression
\begin{equation}
    T_{bh}=\frac{\kappa_{bh}}{2\pi}=\frac{(r_{+}-r_{-})[1-(\alpha r_{+})^2]}{4\pi(r_{+}^2+a^2)}\,,
\end{equation}
in which we regard $a$ still as a free parameter independent of the variance in $r_{+}$, and we write the angular velocity at the outer horizon and extremal one at the acceleration as 
\begin{equation}
    \Omega_{bh}=\frac{\Delta_\phi^{-1}a}{a^2+r_{+}^2}\,,\qquad \Omega_{bh}^{ext}\equiv\tilde{b}=\frac{\Delta_\phi^{-1}a\alpha^2}{1+a^2\alpha^2}\,,
\end{equation}
where $\tilde{b}$ is first introduced in Eq.(\ref{trans}). In the near-extremal limit, the bound states of the Nariai C-metric with boundary condition Eq.(\ref{b-c}) is only regarded as in duality to a chiral half of the 2-D CFT. This corresponds to the fact that between two chiral temperatures one is a finite sum while another precisely vanishes~\cite{Ghosh_2020}, which can also be revealed on gravity side as follows. To compute the chiral temperature, we need to adopt the interpretation of Frolov-Thorne state~\cite{Frolov:1989jh}, which serves as the same role of Hawking-Hartle states in extremal limit, according to Equivalence principle. The energy $\tilde{\epsilon}$ of certain normal modes with angular quantum number $m$ observed by ZAMO observers near the horizon and energy $\epsilon$ observed by distant observers have the match $\tilde{\epsilon}=\epsilon-m\Omega_{H}$, so we have the Bolzman factor rewritten as
\begin{equation}
    e^{-(\omega-m\Omega_H)/{T_{bh}}}=e^{-n_L/T_L-n_R/T_R}
\end{equation}
where $n_L$ and $n_R$ are the quantum numbers corresponding to NHEK coordinates through the transformation in Eq.(\ref{trans}):
\begin{equation}
e^{-i\omega t+im\phi}=e^{-in_R\hat{t}-in_L\hat{\phi}}\,.
\end{equation}
In this way the effective temperature of left moving modes $T_L$ can be calculated as
\begin{equation}
    T_{L}=-\lim_{r_{+}\rightarrow \frac{1}{\alpha}}\frac{T_{bh}}{\Omega_{bh}^{ext}-\Omega_{bh}}=\frac{(1-\alpha r_{-})[1+(a\alpha)^2]}{4\pi a\alpha}\Delta_\phi\,,
\end{equation}
Finally by using the Cardy entropy formula, we have 
\begin{equation}
    S_{\mathrm{CFT}}=\frac{\pi^2}{3}c_L^{(\epsilon)}T_{L}=-\frac{\pi\Delta_{\phi}[1+(a\alpha)^2]}{4\alpha^2}=\frac{1}{4}\mathcal{A}^{(\epsilon)}\,,
\end{equation}
which exactly reproduces Hawking-Bekenstein formula of the black hole entropy.

As discussed in the introduction, this application of the Cardy formula is speculative. There is no conclusive evidence that quantum gravity in a de Sitter background is in fact unitary, given that it only appears as a metastable vacuum in string theory.Tt is more likely that we are calculating something more similar to pseudo entropy rather than entanglement entropy~\cite{Doi_2023}. At the same time, it is not understood how the rotating Nariai geometry
maps to a thermal state in the CFT. Therefore the above formula requires further explanation~\cite{Anninos_2010}.

\subsection{Near extremal-hidden symmetry}
As first pointed out in~\cite{PhysRevD.82.024008}, the scalar field in type-D spacetime with prerequisite that the wavelength of excitation is far larger than the curvature scale, i.e., $\omega m\ll 1$ has a hidden conformal symmetry $SL(2,R)_L\times SL(2,R)_R$, and here we briefly review this idea in the specific case of rotating C-metric in ``Nariai" flat limit. With the metric given by Eq.(\ref{01}), the Klein-Gordon equation for massless charged scalar field $(D_{\mu}D^\mu-\frac{1}{6}R)\Phi=0$, where $D_{\mu}=\partial_{\mu}-iqA_{\mu}$ can be written as
\begin{equation}
    \left\{\partial_r(\Delta\partial_r)+\frac{\Big{[}\frac{am}{\Delta_\phi}-eqr+\omega(a^2+r^2)\Big{]}^2}{\Delta}+\frac{\Delta''}{6}+C\right\}R(r)=0\,,
\end{equation}
in which we have supposed $\Phi=(1-\alpha r\cos\theta)\mathrm{e}^{-i\omega t+ik\phi}\Theta(\theta)R(r)$ and $C$ is a separation constant. Now we consider the following approximating condition:
\begin{enumerate}
    \item Near Nariai limit, and the accelerating horizon $r_s=\frac{1}{\alpha}$ is extremely close to $r_{+}$. So we can approximate $\Delta$ by a quadric function $\Delta\approx \kappa_+(r-r_+)(r-r_s)$, and specifically in Nariai limit we have $\kappa_+=-\frac{2(r_s-r_{-})}{r_s}$.
    \item $\omega r_+/r_s\ll1$, $eq<\approx\omega r_+$. Then we can throw out the residual linear and quadric term of $r$ in the equation and only consider the singular terms. 
\end{enumerate}
With these prerequisites we have 
\begin{equation}
\begin{aligned}
    \{\partial_r((r-r_s)(r-r_+)\partial_r)+\frac{\Big{[}\frac{ak}{\Delta_\phi}-eqr_++\omega(a^2+r_+^2)\Big{]}^2}{\kappa_+^2(r-r_+)(r_+-r_s)}-\frac{\Big{[}\frac{ak}{\Delta_\phi}-eqr_s+\omega(a^2+r_s^2)\Big{]}^2}{\kappa_+^2(r-r_s)(r_+-r_s)}\\
    +O((\omega r)^2)+O(\omega req)+C'\}R(r)=0\,,
\end{aligned}
\end{equation}
in which the first higher order term reads as 
\begin{equation}
    O((\omega r)^2)=[r^2+(r_++r_s)r+r_+^2+r_s^2+r_+r_s+2a^2]\frac{\omega^2}{\kappa_+^2}\,,
\end{equation}
while the second higher order term reads as
\begin{equation}
    O(\omega req)=-\frac{2eq\omega(r+r_++r_s)}{\kappa_+^2}\,,
\end{equation}
and the constant is 
\begin{equation}
    C'=C+\frac{2ak\omega}{\Delta_\phi\kappa_+^2}+\frac{e^2q^2}{\kappa_+^2}\,.
\end{equation}
Now ignoring all higher order terms we can introduce conformal coordinates:
\begin{equation}
    \omega^{+}=\sqrt{\frac{r-r_+}{r-r_s}}\mathrm{e}^{2\pi T_R\phi+2n_Rt}\,,
\end{equation}
\begin{equation}
    \omega^{-}=\sqrt{\frac{r-r_+}{r-r_s}}\mathrm{e}^{2\pi T_L\phi+2n_Lt}\,,
\end{equation}
\begin{equation}
    y=\sqrt{\frac{r_+-r_s}{r-r_s}}\mathrm{e}^{\pi (T_R+T_L)\phi+(n_R+n_L)t}\,,
\end{equation}
but as the common case with nonzero charges need to take $Q$ picture which is typically ill-defined for non-extremal black hole~\cite{}, we have to take $q=0$ here and take $J$ picture, leading to the following results:
\begin{equation}
    T_R=\frac{\kappa_+(r_+-r_s)\Delta_\phi}{4\pi a}\,,\quad T_L=\frac{\kappa_+(r_+^2+r_s^2+2a^2)}{4\pi a(r_++r_s)}\,,\quad n_R=0\,,\quad n_L=-\frac{\kappa_+}{2(r_++r_s)}\,.
\end{equation}
and then by defining
\begin{equation}
    H_+=i\frac{\partial}{\partial \omega^+}\,,\quad
    H_{0}=i\Big{(}\frac{\partial}{\partial\omega^+}+\frac{y}{2}\frac{\partial}{\partial y}\Big{)}\,,\quad
    H_{-}=i\Big{(}(\omega^+)^2\frac{\partial}{\partial\omega^+}+\omega^+y\frac{\partial}{\partial y}-y^2\frac{\partial}{\partial\omega^-}\Big{)}\,,
\end{equation}
\begin{equation}
    \Bar{H}_+=i\frac{\partial}{\partial \omega^-}\,,\quad
    \Bar{H}_{0}=i\Big{(}\frac{\partial}{\partial\omega^-}+\frac{y}{2}\frac{\partial}{\partial y}\Big{)}\,,\quad
    \Bar{H}_{-}=i\Big{(}(\omega^-)^2\frac{\partial}{\partial\omega^-}+\omega^+y\frac{\partial}{\partial y}-y^2\frac{\partial}{\partial\omega^+}\Big{)}\,,
\end{equation}
one can easily find these operators having $sl(2,R)\times sl(2,R)$ algebra:
\begin{equation}
    [H_0,H_{\pm}]=\mp iH_{\pm}\,,\qquad [H_-,H_+]=-2iH_0\,,
\end{equation}
\begin{equation}
    [\Bar{H}_0,\Bar{H}_{\pm}]=\mp i\Bar{H}_{\pm}\,,\qquad [\Bar{H}_-,\Bar{H}_+]=-2i\Bar{H}_0\,,
\end{equation}
and the Laplace operator for scalar field turns out to be the Casimir operator of the algebra:
\begin{equation}
    \mathbf{H}^2=-H_0^2+\frac{1}{2}(H_+H_-+H_+H_-)\,.
\end{equation}

\section{Going from extremal to near extremal-Reduction to JT gravity model}\label{sec:5}
Before moving on to the JT reduction, we first present the connection of the reduction we apply below to the holography duality of the Sec.\ref{sec:4}. The JT type action is often used in presenting the $n \mathrm{AdS_2}/n\mathrm{CFT_1}$ correspondence, which, as another perspective of view, applies to the near-horizon region of a black hole that is nearly extremal. In 4D the conventional Near Horizon Extremal
Kerr (NHEK) limit that forms the basis for the Kerr/CFT correspondence is interpreted as a trivial IR fixed point of the dual field theory of JT gravity, which corresponds to a saddle point of it in which all degrees of freedom take their attractor value. Inversely it is the extension of the geometry away from NHEK that adds dynamics to the theory by irrelevant operators in field theory side, which provides the breaking from strict $\mathrm{AdS}_2/\mathrm{CFT}_1$ to $n \mathrm{AdS_2}/n\mathrm{CFT_1}$~\cite{Castro_2018}. Generally through JT reduction the scale breaking effect of the dual theory will be precisely captured by dilation field $\Phi$. Moreover, the dynamical modes we discuss in great detail in Kerr/CFT (ASG) are viewed as those forming a solution with ``hair" away from classical solution; they are ``large diffeomorphic" to the latter, which can alter the boundary condition we impose on the dilation and thus serve as Goldstone modes in JT gravity framework.

The similarity and connection between extremal charged black hole and JT gravity model is well-known and intensively studied. Unlike spherically symmetric solutions, rotating black holes have much more complicated modes excitations originating from different metric components, even in extremal case. However, a relatively simple near-horizon form can lead to a typical $\mathrm{(A)dS_2}$ geometry, and the transverse volume expanded by two angular coordinates can be represented by something proportional to the dilation. The solution obtained in this circumstance can be regarded as an attractor value of a geneal solution family, thus the deviation from extremal to non extremal can only lead to variation of the action only up to the second order, so when aiming to study the leading order thermodynamic behavior, we only need to study linear variations excited from the attractor value, and the form of the action is still JT-type. This treatment of course restricts the scope of application of this approximation, so we have to divide the whole Poincare patch by two regions, one near-horizon and one far into the conformal boundary. These two regions are separated by $\partial_{\mathrm{(A)dS_2}}$, and it can be proven that the JT gravity model restricted to the near-horizon region is correctly equivalent to the whole 4-D theory.

Rewrite the metric Eq.(\ref{01}) into the form 
\begin{equation}
    \mathrm{d}s^2=\frac{1}{\Omega^2}[-\frac{\Delta\rho^2}{\Sigma}\mathrm{d}t^2+\frac{\rho^2}{\Delta}\mathrm{d}r^2+\Phi^2\frac{\rho^2}{P\sqrt{\Sigma}}\mathrm{d}\theta^2+\Phi^2\frac{\sin^2\theta P\sqrt{\Sigma}}{\rho^2}(\Delta_\phi\mathrm{d}\phi-\omega\mathrm{d}t)^2]\,,
\end{equation}
where 
\begin{equation}\label{phi}
    \omega=\frac{Pa(r^2+a^2)-\Delta a}{P(r^2+a^2)^2-a^2\sin^2\theta\Delta}\,, \qquad 
    \Phi^2=\sqrt{\Sigma}=\sqrt{(r^2+a^2)^2-\frac{a^2\Delta}{P}\sin^2\theta}\,,
\end{equation}
the reason for choosing this parametrization is that the volume of the internal two sphere spanned by $\theta, \phi$ is now
given by $\Phi^2\int\mathrm{d}S \frac{1}{\Omega^2}$ and therefore only dependent on $\Phi$ and manifestly independent of $\Sigma$. On the other hand, in the extremal case, $\Phi^2$ does not depend on $\theta$ and thus can be regard as a function of $t$ and $r$. We regard the dilation of the two dimension spacetime as originating from the fluctuation of $\Phi^2$ near its attractor value, which makes sense when the black hole is extremal, and the specific behavior of the fluctuation is crucial for thermodynamics.

Now for extremal case, first consider the gravitation action of the general form in 4 dimension~\cite{Moitra_2019}
\begin{equation}\label{i4}
    I_{G}=-\frac{1}{16\pi G_{4}}\int\mathrm{d}^4x\sqrt{-g}R-\frac{1}{8\pi G_4}\int_{\mathcal{B}}\sqrt{\gamma}K\,,
\end{equation}
in which $G_{4}$ is the 4-D Newton constant, $\mathcal{B}$ is the boundary of Poincare patch, $\gamma$ and $K$ is the intrinsic metric and extrinsic curvature. Now suppose the metric to be
\begin{equation}\label{ans}
\mathrm{d}s^2=f(\theta)g_{ab}\mathrm{d}x^a\mathrm{d}x^b+h(\theta)\Phi^2\mathrm{d}\theta^2+p(\theta)\Phi^2(\mathrm{d}\phi+A_{a}\mathrm{d}x^a)^2\,,
\end{equation}
where $a,b$ stands for coordinates $t,r$ and $\Phi^2$ is regarded as the dilation, while $A_{a}$ stands for the gauge field, which merely depends on $t,r$. Regarding that $\Delta_\phi$ is an independent parameter, for simplicity we here take it as 1. Follow the most common Kaluza-Klein dimension reduction procedure, (for reduction detail of similar process see~\cite{Moitra_2019,heydeman2021statisticalmechanicsnearbpsblack,Iliesiu_2021}) we are able to express Eq.(\ref{i4}) as
\begin{equation}
\begin{aligned}
    I_{G}=-\frac{1}{8G_{4}}\int_{0}^{\pi}\mathrm{d}\theta\int\mathrm{d}^2x\sqrt{-g_{2}}\Big{[}\sqrt{hp}(\Phi^2R_{2}-4\Phi\nabla^2\Phi-2(\nabla\Phi)^2)-\frac{1}{4}\Phi^4\frac{\sqrt{hp^3}}{f}F_{ab}F^{ab}\qquad\qquad\\
   +\sqrt{\frac{p}{h}}\Big{(}\frac{f(h'p'-2hp'')}{2hp}-\frac{p'f'}{p}+\frac{fp'^2}{2p^2}+\frac{f'^2}{2f}+\frac{h'f'-2hf''}{h}\Big{)}\Big{]}
    -\frac{1}{4G_4}\int_0^{\pi}\mathrm{d}\theta\int\mathrm{d}t\sqrt{\gamma_2}\sqrt{fhp}\Phi^2K_2\,.\qquad\qquad
\end{aligned}
\end{equation}
where $g_{2},\gamma_2$ stands for the determinant of the 2-D metric of $g_{ab}$ and its induced metric $\gamma_{ab}$ on $r=\infty$ respectively, $R_{2}$ stands for the 2-D scalar curvature, and $K_2$ is the extrinsic curvature of $r=\infty$. We have thrown out sums that is nonzero when $g_{ab}$ is flat in the boundary term. At the same time, we should not forget the contribution of the EM action. If we suppose the vector potential of the form
\begin{equation}
\Tilde{A}=k(\theta)A_a\mathrm{d}x^a+b(\theta)\mathrm{d}\phi\,,
\end{equation}
then EM action $I_{EM}=\frac{1}{16\pi G_4}\int\mathrm{d}^4xF^2$ can be reduced to
\begin{equation}
\begin{aligned}\label{em}
I_{EM}=\frac{1}{8G_4}\int_0^\pi\mathrm{d}\theta\int\mathrm{d}^2x\sqrt{-g}\Big{[}\frac{\sqrt{hp}}{f}k^2\Phi^2F_{ab}F^{ab}+2(k'-b')^2\sqrt{\frac{p}{h}}A_{a}A^a
+2\frac{f}{\sqrt{hp}}\frac{b'^2}{\Phi^2}\Big{]}\\
+\mathrm{boundary\; terms\; related\; to \;phase\; modes}\,.
\end{aligned}
\end{equation}

Now we focus on the reduced metric and abandon these indices in following calculation. When the rotating C-metric is in the usual extremal limit, i.e., $r_{+}=r_{-}=r_{e}$, and
\begin{equation}\label{fhp}
    f(\theta)=\frac{1+u^2\cos^2\theta}{2(1-\alpha r_e\cos\theta)^2}\,,\qquad 
    h(\theta)=\frac{1+u^2\cos^2\theta}{(1-\alpha r_e\cos\theta)^4(1+u^2)}\,,\qquad
    p(\theta)=\frac{\sin^2\theta(1+u^2)}{(1+u^2\cos^2\theta)}\,,
\end{equation}
in which $u=\frac{a}{r_e}$, and we have
\begin{equation}\label{ext}
\begin{aligned}
    I_{G}=-\frac{1}{4G_{4}}\int\mathrm{d}^2x\sqrt{-g}\Big{[}\frac{\Phi^2R-4\Phi\nabla^2\Phi-2(\nabla\Phi)^2}{1-\alpha^2r_e^2}-\frac{\Phi^4}{2}\frac{1+u^2}{2}\frac{u+(u^2-1)\mathrm{arctan}u}{u^3}F_{ab}F^{ab}\\
    +(1+u^2)\frac{u+(1-u^2)\mathrm{arctan}u}{2u}\Big{]}-\frac{1}{4\sqrt{2}G_4}\int\mathrm{d}t\sqrt{\gamma}g(\alpha r_e,u)\Phi^2K\,,\qquad\qquad\qquad
\end{aligned}
\end{equation}
where
\begin{equation}
\begin{aligned}
    g(x,y)=\frac{1}{2}\Big{[}\frac{2y^2(x^2+1)+4x^2}{(x^2-1)^2}\frac{\sqrt{1+y^2}}{x^2+y^2}\qquad\qquad\qquad\\
    +\frac{y^2}{(x^2+y^2)^{3/2}}\Big{(}\mathrm{ln}\Big{(}\frac{1+x}{1-x}\Big{)}+\mathrm{ln}\Big{(}\frac{x+y^2+\sqrt{1+y^2}\sqrt{x^2+y^2}}{x-y^2+\sqrt{1+y^2}\sqrt{x^2+y^2}}\Big{)}\Big{)}\Big{]}\,,
\end{aligned}
\end{equation}
and the attractor value corresponds to
\begin{equation}
    \Phi_{0}^2=r_{e}^2+a^2\,,\quad g_{ab}\mathrm{d}x^a\mathrm{d}x^b=\frac{2r_{e}^2}{1-\alpha^2r_e^2}\Big{(}-r^2\mathrm{d}t^2+\frac{\mathrm{d}r^2}{r^2}\Big{)}\,,\quad
    A_{a}=\frac{2ar_er}{(1-\alpha^2r_{e}^2)(a^2+r_e^2)}\delta_{at}\,,
\end{equation}
In extremal limit, the gauge field is described by
\begin{equation}\label{kk}
    k(\theta)=-e\frac{u^2+1}{2u}\frac{1-u^2\cos^2\theta}{1+u^2\cos^2\theta}\,,\qquad b(\theta)=-e\frac{u\sin^2\theta}{1+u^2\cos^2\theta}\,,
\end{equation}
and bring them into Eq.(\ref{em}) in we are able to get
\begin{equation}
\begin{aligned}
    I_{EM}=\frac{e^2(1+u^2)^2}{8G_4}\int\mathrm{d}^2x\sqrt{-g}\frac{1}{2u^2}\Big{[}\frac{1-u^2}{(1+u^2)^2}+\frac{\mathrm{arctan}u}{u}\Big{]}\Phi^2F_{ab}F^{ab}\\
    +\frac{1}{\Phi^2}\Big{[}\frac{u^2-1}{(1+u^2)^2}+\frac{\mathrm{arctan}u}{u}\Big{]}\,,
\end{aligned}
\end{equation}
as we see, the crucial relation $k(\theta)-b(\theta)=-e\frac{1-u^2}{2u}$ guarantees that the reduced EM field is still massless. Here, parameter $e$ is not independent given two action parameters and static solution, but can be expressed as $\frac{\Phi_0^2}{2}(1-u^2)$. Take the limit of $u\rightarrow 1$ and $\alpha\rightarrow 0$, we get exactly the same result as in~\cite{Moitra_2019} for the reduction of the Kerr BH.

In ``Nariai" flat limit 
\begin{equation}
    f(\theta)=\frac{1+\alpha^2 a^2\cos^2\theta}{2(1-\cos\theta)^2}\,,
\end{equation}
\begin{equation}
    h(\theta)=\frac{1+\alpha^2a^2\cos^2\theta}{(1-\cos\theta)^3(1-\alpha r_{-}\cos\theta)(1+\alpha^2a^2)}\,,
\end{equation}
\begin{equation}
     p(\theta)=\frac{\sin^2\theta(1-\alpha r_{-}\cos\theta)(1+a^2\alpha^2)}{(1-\cos\theta)(1+\alpha^2a^2\cos^2\theta)}\,,
\end{equation}
we have the regularized action (here we regularize in the scheme of a fixed $\theta=\delta$). 
\begin{equation}\label{nar}
\begin{aligned}
    I^{(\delta)}=\frac{1}{8G_4}\int\mathrm{d}^2x\sqrt{g}[\frac{1}{6}(\Phi^2R-4\Phi\nabla^2\Phi-2(\nabla\Phi)^2)+\frac{\Phi^4}{4}g_1(\alpha a,\alpha r_{-})F_{ab}F^{ab}+g_2(\alpha a,\alpha r_{-})]\\
    -\frac{1+a^2\alpha^2}{4G_4}\int\mathrm{d}t\frac{5\alpha^2a^2-1}{3\sqrt{1+\alpha^2a^2}}\sqrt{\gamma}K\,,\qquad\qquad\qquad
\end{aligned}
\end{equation}
where 
\begin{equation}
    g_{1}(x,y)=2\frac{x^3+xy+(1+x^2)(x^2-y)\mathrm{arctan}x}{x^3}\,,
\end{equation}
\begin{equation}
\begin{aligned}
    g_{2}(x,y)=\Big{(}\frac{2}{3}(y-1)+2(x^2-y)\frac{\mathrm{arctan}x}{x}\Big{)}\frac{1+x^2}{2}+x^2+y\,,
\end{aligned}
\end{equation}
and the attractor value is 
\begin{equation}
    \Phi_{0}^2=\frac{1}{\alpha^2}+a^2\,,\quad g_{ab}\mathrm{d}x^a\mathrm{d}x^b=\frac{1}{\alpha^2(1-\alpha r_{-})}\Big{(}r^2\mathrm{d}t^2-\frac{\mathrm{d}r^2}{r^2}\Big{)}\,,\quad
    A_{a}=\frac{ra\alpha}{(1-\alpha r_{-})(1+a^2\alpha^2)}\delta_{at}\,.
\end{equation}
Meanwhile, the gauge field is given by
\begin{equation}
    k(\theta)=-\frac{e(1-\alpha^2a^2\cos^2\theta)(1+\alpha^2a^2)}{2(1+\alpha^2a^2\cos^2\theta)a\alpha}\,,\qquad
    b(\theta)=-\frac{ea\alpha\sin^2\theta}{1+a^2\alpha^2\cos^2\theta}\,,
\end{equation}
so the reduced EM action is
\begin{equation}
\begin{aligned}
    I_{EM}=\frac{e^2(1+a^2\alpha^2)}{16G_4}\int\mathrm{d}^2x\frac{(1+a^2\alpha^2)^2}{2a^2\alpha^2}\Big{[}\frac{\mathrm{arctan}a\alpha}{a\alpha}+\frac{1-a^2\alpha^2}{(1+a^2\alpha^2)^2}\Big{]}\Phi^2F_{ab}F^{ab}\\
    +\frac{4}{\Phi^2}\Big{[}\frac{\mathrm{arctan}a\alpha}{a\alpha}+\frac{a^2\alpha^2-1}{(1+a^2\alpha^2)^2}{}\Big{]}\,.
\end{aligned}
\end{equation}

Further processing needs us to make the transformation
\begin{equation}\label{jt1}
    \Phi=\Phi_0(1+\phi)\,, \qquad g_{ab}\rightarrow g_{ab}\frac{\Phi_{0}}{\Phi}\,,
\end{equation}
in the near-horizon region, where $\phi$ stands for usual definition of dilation field and should be distinguished from the angular coordinate introduced in previous chapters. The attractor value we discuss above, which corresponds to constant dilation scalars are IR fixed point of the equation of motion as defined in~\cite{Castro_2018}. For fluctuations, we only keep up to the first order perturbation of the field $\phi$, and then the bulk term of Eq.(\ref{ext}) (usual extremal action) turns into 
\begin{equation}\label{expre}
I_{\mathrm{JT}}=\mathrm{ground\;terms}+\frac{3a_1\Phi_0^2}{4G_4}\int_{\partial_{\mathrm{AdS_2}}}\mathrm{d}x\sqrt{\gamma}n^a\nabla_a\phi-\frac{\Phi_0^2}{4G_4}\int\mathrm{d}^2x\sqrt{g}[a_1\phi (R-\Lambda)+a_2\phi\Phi_0^2F_{ab}F^{ab}]\,,
\end{equation}
where 
\begin{equation}
a_1=\frac{2}{1-\alpha^2r_e^2}\,,
\end{equation}
\begin{equation}
\Lambda=\frac{1}{a_1\Phi_0^2}\Big{[}1-2\frac{1-u^2}{u}\mathrm{arctan}u+3\Big{(}\frac{1-u^2}{1+u^2}\Big{)}\Big{]}\,, 
\end{equation}
\begin{equation}
    a_2=\frac{1+u^2}{2}\frac{(1-u^2)\mathrm{arctan}u}{u^3}-\frac{5(1+u^2)^2+3(1-u^2)^2}{4u^2(1+u^2)}\,,
\end{equation}
and for Nariai case things are rather similar, which also accords to the general form of JT gravity model. 

Now we can verify that the reduction to 2D provides a consistent description of on-shell dynamics in 4D,i.e., ant solution that satisfies Einstein equation in the ansatz in Eq.(\ref{ans}) is also a solution of JT action and vise versa. As usual, when considering the on-shell action, we can integrate out the dilation configuration and get the equation of motion for metric:
\begin{equation}\label{eom1}
    a_1(R-\Lambda)+a_2\Phi_0^2F_{ab}F^{ab}=0\,,
\end{equation}
At the same time with variation of the EM field we can get the Maxwell equation 
\begin{equation}\label{eom2}
    \partial_{\mu}(\sqrt{g}\phi F^{\mu\nu})=0\,,
\end{equation}
 as expected. With the variation of the metric we can get the EOM of dilation:
\begin{equation}\label{eom3}
    a_1(\nabla_{a}\nabla_b\phi-g_{ab}\nabla^2\phi-g_{ab}\frac{\Lambda}{2}\phi)=2a_2\Phi_0^2\phi(F_{ac}F_{b}^{\,c}-\frac{1}{4}g_{ab}F_{cd}F^{cd})\,,
\end{equation}
after this procedure by adding the counter term to regularize, we have the boundary term as a Schwarzian action (proportional to $\mathrm{Sch}\left\{t,\tau\right\}\equiv\frac{t'''}{t'}-\frac{3}{2}(\frac{t''}{t'})^2$), whose contribution comes from re-parameterization of the conformal boundary (Schwarzian modes). As the rest process is much like the common treatment of JT gravity~\cite{Mertens_2023,Moitra_2019,Lam_2018,Yang_2019,saad2019latetimecorrelationfunctions,jafferis2023jtgravitymattergeneralized,harlow2019factorizationproblemjackiwteitelboimgravity,Brown_2019}, we omit them here. 

We now present more physical interpretation of the above reduction. The reduction to JT gravity model, made by Eq.(\ref{jt1}), is only rational for the region between the horizon ($r=0$) and the $\mathrm{(A)dS_2}$
 boundary $\partial_{\mathrm{(A)dS_2}}$, which satisfies the condition that the effects of finite temperature have died down, but not so far that the effects of the breaking of scale invariance have become significant. If we denote temperature as $T$, and energy scale to measure the breaking as $\mathcal{J}$, the region is where $T\ll r^{-1}\ll \mathcal{J}$. If we divide the whole action in previous calculation into two parts: $I=I_{[H\rightarrow \partial_{\mathrm{(A)dS_2}}]}+I_{[\partial_{\mathrm{(A)dS_2}}\rightarrow\infty]}$, then it is proven that~\cite{Moitra_2019}
\begin{equation}
\delta I_{[H\rightarrow \partial_{\mathrm{(A)dS_2}}]}=\delta I_{\mathrm{JT}}^{\mathrm{bulk}}\,,
\end{equation}
\begin{equation}
\delta I_{[\partial_{\mathrm{(A)dS_2}}\rightarrow\infty]}=\delta I_{\mathrm{JT}}^{\mathrm{boundary}}\,    
\end{equation}
to the leading order, which means that the JT gravity model correctly reflects the whole thermodynamics. Explicitly, we first compute the total ADM energy/free energy in JT gravity theory reduced from usual extremal case. Suppose the characteristic scale to break the conformal invariance to be $\mathcal{J}$, i.e., we have
\begin{equation}
    \phi=\frac{1}{\mathcal{J}z}+O(z)\,, \qquad z\rightarrow 0\,,
\end{equation}
where $\frac{1}{z}=\frac{(r-r_e)(1-\alpha^2r_e^2)}{r_e^2+a^2}$, and $1/\mathcal{J}$ has the dimension of length. Then the (ADM) energy is given by~\cite{Mertens_2023}
\begin{equation}
    E_{JT}=\frac{\pi}{4\tilde{G}\mathcal{J}}T^2=-\Delta F\,,
\end{equation}
here $\Delta F$ is the free energy deviated from the extremal value, $\tilde{G}$ is the coupling constant for JT theory. By expanding Eq.(\ref{phi}) near $r=r_e$ we have $1/\mathcal{J}=r_e/(1-\alpha^2r_e^2)$, considering that $\Delta$ has a double root near the extremity. This $z$-expansion scheme produces a $\mathcal{J}$ satisfying its original definition
\begin{equation}
    \Delta S= \frac{\pi}{2\tilde{G}}\frac{T}{\mathcal{J}}\,,
\end{equation}
for the entropy above extremity in JT gravity.  
So we have $E_{JT}=\frac{\pi r_eT^2}{4\tilde{G}(1-\alpha^2r_e^2)}$. With the relation between the JT coupling constant $\tilde{G}$ and coupling constant of the full theory $G_4$ (one can confirm this directly from the coefficient of the last term in Eq.(\ref{expre}))
\begin{equation}
    \frac{1}{\tilde{G}}=\frac{1}{G_4}\frac{8\pi\Phi_0^2}{1-\alpha^2r_e^2}\,,
\end{equation}
we have the total energy of the 4D theory 
\begin{equation}
    E_{tot}=\frac{2\pi^2r_e(r_e^2+a^2)T^2}{G_4(1-\alpha^2r_e^2)^2}\,,
\end{equation}
and this is consistent with the expansion of the mass parameter (pure ADM, not the integrable mass we construct before) near extremity in canonical ensemble (fixing the total charge $J$, $Q$ and acceleration $\alpha$):
\begin{equation}
r_{+}=r_e+\frac{2\pi(r_e^2+a^2)}{1-\alpha^2r_e^2}T+\frac{2\pi^2(r_e^2+a^2)r_e(5-\alpha^2r_e^2+4\alpha^2a^2)}{(1-\alpha^2r_e^2)^3}T^2\,,
\end{equation}
\begin{equation}
r_{-}=r_e-\frac{2\pi(r_e^2+a^2)}{1-\alpha^2r_e^2}T-\frac{2\pi^2(r_e^2+a^2)r_e(3+\alpha^2r_e^2+4\alpha^2a^2)}{(1-\alpha^2r_e^2)^3}T^2\,,
\end{equation}
(where $a$ represents for the extremal value), and 
\begin{equation}
    M=r_e+2\pi^2\frac{r_e(r_e^2+a^2)}{(1-\alpha^2r_e^2)^2}T^2\,,
\end{equation}
given that $E_{tot}=M/G_4$.

\section{Conclusion and discussion}\label{sec:6}
In this study, our first step is to present the first law of thermodynamics of the rotating C-metric. Many difficulties have been found when intending to construct the first law of rotating C-metric. Although there have been many attempts in this realm, either the result is based on parameter perturbation and integrability analyses, which lacks geometrical explanation, or we have the awkward result that the mass of C-metric is zero. The main problems include the treatment of the infinite area of acceleration horizon and the existence of conical singularity, accompanied with cosmic string.  The first one leads to the debate on whether the mass defined by us should indeed contain contribution from the acceleration horizon or not. The second one, on the other hand, the variation of the cosmic string can also contribute to the first law, and only by some redefinition of the mass we can reproduce integrability. This needs the regularization of the area and also, Komar integral.

We also reveal the holography duality between the rotating C-metric in ``Nariai" flat limit and warped $\mathrm{CFT}_2$, and reduce the action to 2 dimension in order to find the correspondence of the extremal black hole to JT gravity. Based on the analyses of the thermodynamic variables, we can confidently handle the holography duality when the event horizon and the acceleration horizon coincide, which is exactly the Nariai limit of the rotating C-metric. We find the results are still as expected: the results of the entropy obtained by two dual aspects finally agree with each other, which again proves the correctness of Nariai-CFT correspondence. Still, because there is no existing self-consistent theory to describe the quantum gravity in spacetime with positive cosmological constant, we still lack specific details in presenting this holography dual, and the deeper reason for the results to occur is still unclear. All these problems require profound thoughts in the future. Finally, because there have been a large amount of interesting properties contained in JT gravity system, we present the reduction of the action to the JT form for both the usual extremal limit and Nariai limit of the rotating C-metric, which can provide the basis to many well-known treatments to investigate further quantum effects in this frame. 
\section{acknowledgments}
The author is grateful for discussions with Xin Meng and Yu-Sen Zhou.
\section{Declarations}
The author declare no competing interests.
\bibliography{reference}{}
\bibliographystyle{apsrev4-1}
\end{document}